\documentclass[
reprint,
 superscriptaddress,
 showpacs,preprintnumbers,
 amsmath,amssymb,
 aps,
pra,
 floatfix,
]{revtex4-1}

\usepackage[T1]{fontenc} 
\usepackage{mathptmx}

\usepackage{graphicx}
\usepackage{dcolumn}
\usepackage{bm}
\usepackage[colorlinks,
breaklinks,
 citecolor=blue,linkcolor=blue,urlcolor=blue
]{hyperref}

\begin{document}

\title{Laser cooling with adiabatic passage for diatomic molecules}%

\author{Qian Liang}

\author{Tao Chen}%
\email{phytch@zju.edu.cn}

\author{Wenhao Bu}

\author{Yuhe Zhang}
\affiliation{%
Zhejiang Province Key Laboratory of Quantum Technology and Device, Department of Physics 
and State Key Laboratory of Modern Optical Instrumentation, Zhejiang University, Hangzhou, Zhejiang, China, 310027
}%
\author{Bo Yan}
\email{yanbohang@zju.edu.cn}
\affiliation{%
Zhejiang Province Key Laboratory of Quantum Technology and Device, Department of Physics 
and State Key Laboratory of Modern Optical Instrumentation, Zhejiang University, Hangzhou, Zhejiang, China, 310027
}%
\affiliation{%
 Collaborative Innovation Centre of Advanced Microstructures, Nanjing University, Nanjing, China, 210093
}%

\date{\today}

\begin{abstract}

We present a magnetically enhanced laser cooling scheme applicable to multi-level type-II transitions
and further diatomic molecules with adiabatic transfer. An angled magnetic field is introduced to not
only remix the dark states, but also decompose the multi-level system into several two-level sub-systems in
time-ordering, hence allowing multiple photon momentum transfer. For complex multi-level diatomic molecules, although
the enhancement gets weakened, our simulations still predict a $\sim 4\times$ larger value of the maximum
achievable cooling force and a wider coolable velocity range compared to the conventional Doppler cooling.
A reduced dependence on spontaneous emission of this scheme makes laser cooling a molecule with
leakage channels become a feasibility.

\end{abstract}
\maketitle

\section{Introduction}

Allowing precise manipulation and fast preparation of cold atomic samples \cite{Chu1998,Phillips1998,Cohen-Tannoudji1998},
laser cooling has revolutionized the development of atomic, molecular, optical and quantum physics during last several decades
\cite{Cornell2002,Bloch2008,Ludlow2015,Bohn2017,Safronova2018}. Conventional Doppler
cooling only requires a cooling laser with one single frequency, typically red-detuned. It is good enough for
almost all atomic species. But for some special cases, lots of multi-frequency cooling schemes \cite{Metcalf2017}, including adiabatic rapid passage \cite{
Lu2005, Miao2007, Jayich2014} and bichromatic force \cite{Soeding1997,Yatsenko2004,Partlow2004,Corder2015}, have been
proposed and predicted to have a better performance but with the price of increasing the complexity of the system. 
However, some features, for example, stronger cooling forces and a weak dependence on 
spontaneous emission \cite{Metcalf2017}, make them potentially be applied in direct laser cooling molecules 
where the cycling transition is quasi-closed \cite{Shuman2010,Hummon2013,DiRosa2004,Chen2016}. Besides the leakage channels, the 
energy levels for molecules are complex and type-II transitions dominate \cite{Stuhl2008,Chen2017}, making the cooling process much
more complex and the Doppler cooling force much weaker than those in atomic cases. To achieve a better cooling efficiency, it might be worthwhile
to sacrifice the simplicity of the cooling scheme by introducing a light field with multiple frequencies \cite{Kozyryev2018}.

To describe the mechanism of laser cooling, an atom-light interaction approach in either semi-classical or quantum picture can be 
directly applied to explain the momentum transfer \cite{Metcalf1999, DALIBARD1989, UNGAR1989}. The strength of the radiation force
depends on how fast the momentum exchange is repeated, that is, the decay rate from the excited state. Once 
successive scattering was achieved in a closed cycling transition, the force would only be limited by the spontaneous decay rate $\Gamma$.
For a two-level system, the maximum force $F_\text{rad,max}=\hbar k\Gamma/2$ with $\hbar k$ the photon momentum. The cooling
velocity range and the temperature limit also depend on the decay rate $\Gamma$ \cite{Metcalf1999}.

\begin{figure*}[]
\includegraphics[width=0.95\textwidth]{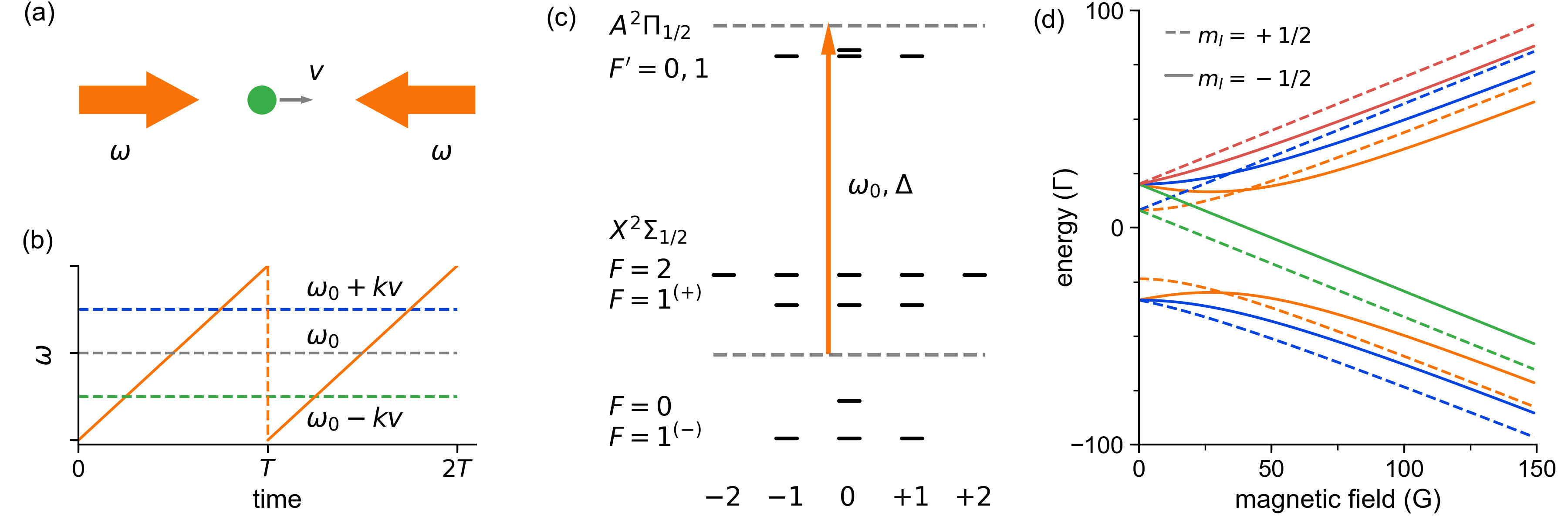}
\caption{\label{fig1} (Color online) (a) Laser cooling scheme with sawtooth
wave adiabatic passage. The two counter-propagating laser beams with a time-dependent frequency
of $\omega$ interact with a moving particle with a velocity of $v$.
(b) The laser frequency $\omega$ linearly ramps in one sweeping period
$T$. Due to Doppler effect, the two beams become resonant with the
moving particle at different times indicated by the green (counter-propagating) and blue (co-propagating) dashed lines.
The gray dashed line labels the center frequency $\omega_0$ of the sweeping.
(c) Cycling transition 
for typically laser-coolable diatomic molecules under a weak magnetic field, only rotational,
hyperfine and Zeeman branches are shown. The energy gap between the two dashed lines
indicates the center frequency $\omega_0$ of the chirped laser beams, and
$\Delta$ is the energy shift of the center frequency corresponding to the resonant condition ($\omega_0, \Delta=0$) defined by 
the energy gap between the zero position of the four ground hyperfine states in (d) and the upper $F'=1$ state.
(d) Energy splittings of the $|X,N=1,-\rangle$ ground state of the BaF molecule under external magnetic field.
The hyperfine states are labelled with $|S,m_S;N,m_N; I, m_I\rangle$. $m_I=+1/2$ and $m_I=-1/2$ are plotted with
dashed and solid lines respectively, while different values of $m_S$ and $m_N$ are shown in four different colors: $m_S=-1/2,~m_N=-1$ (green),
$m_S=-1/2,~m_N=0$ (orange, lower), $m_S=-1/2,~m_N=+1$ (blue, lower) and $m_S=+1/2,~m_N=+1$ (red), $m_S=+1/2,~m_N=0$ (blue, upper),
$m_S=+1/2,~m_N=-1$ (orange, upper). 
The lower four states belong to $J=1/2$ branch while others are 
in $J=3/2$ branch. }
\end{figure*}

In order to go beyond the limit of the spontaneous emission, the stimulated emission has been introduced to perform laser cooling.
Recently, an adiabatic-passage cooling scheme by rapidly sweeping the laser frequency in a sawtooth wave shape (SWAP) has been 
experimentally demonstrated in narrow-line transitions \cite{Norcia2018,Muniz2018,Petersen2018} and Raman transitions \cite{Greve2018}.
Such a scheme uses the stimulated emission to transfer the excited atoms back to the ground state, therefore strong forces can be achieved
with a high sweeping repetition rate even for transitions with a small $\Gamma$. Let us consider a moving particle with a velocity
of $v$ in a light field from two counter-propagating laser beams with a time-dependent frequency $\omega$; see Fig.\ref{fig1}(a). Due
to the Doppler effect, resonances of the particle with the two beams are separated in time ordering. With an increasing ramp of the frequency
in Fig.\ref{fig1}(b), an adiabatic excitation is first induced by the counter-propagating beam, followed by another adiabatic deexcitation 
process triggered by the co-propagating beam. Ideally, the particle loses twice the photon momentum during one sweeping period, resulting in
a maximum average force of $F_\text{swap}=2\hbar k T^{-1}$. To guarantee the adiabatic transfer, the Landau-Zener condition $\Omega^2\gg\alpha$
should be fulfilled \cite{Bartolotta2018}, here $\Omega$ is the on-resonance Rabi frequency and $\alpha$ is the frequency ramp speed.

Now an interesting question is whether the SWAP scheme can be extended to enhance the cooling effect for multi-level type-II transitions 
and further complex molecules. The multi-level type-II transitions generally show a weak Doppler cooling force \cite{Oien1997,Tiwari2008} 
due to the remixing process of the dark states via either rapidly switching the polarization of the light \cite{Hummon2013,Anderegg2017} 
or introducing an angled magnetic field to re-define the quantum axis \cite{Shuman2010,Truppe2017}. Here we consider the latter case. Distinct from the
ideal two-level system where the excitation from counter-propagating beam strictly followed by stimulated deexcitation from co-propagating beam,
the problem of the multi-level systems lies in that the excitation and stimulated deexcitation pairs might become out of
order, that is, the roles of the three polarization components in the two beams get mixed even the resonant positions of two sublevels are
separated due to the energy gap introduced by a magnetic field. This will necessarily 
degrade the quality of the momentum transfer. Diatomic molecules that can be directly laser cooled generally have a 4+12 level structure, as illustrated in
Fig.\ref{fig1}(c) [under weak magnetic field]. It makes the momentum transfer much more complex. In this work, we apply a time-dependent master
equation approach to describe the interaction between a multi-level system and a frequency chirped light field in Sec.\ref{sec2}, and then we investigate 
the SWAP cooling effect for type-II transitions and diatomic molecules under both weak and strong magnetic field regimes in Sec.\ref{sec3} and Sec.\ref{sec4}. 
Finally, in Sec.\ref{sec5}, we consider a three-level leakage system to check the resistance of the SWAP scheme on the unwanted loss channels. 

\section{Time-dependent master equation approach}\label{sec2}

Without loss of generality, we consider the interaction between a multi-level particle and a multi-frequency laser field. An arbitrary multi-frequency 
light field can be decomposed as
\begin{equation}\label{eq1}
 \vec{E} = \sum\limits_{p,q} \frac{1}{2}E_p M_{pq}
 \hat{\epsilon}_q e^{i\vec{k}_p\cdot\vec{r}} e^{-i\omega_p t} + \text{c.c.} ,
\end{equation}
where $p$ is the laser beam index with frequency $\omega_p$ and wave vector (propagating direction) $\vec{k}_p$, and the amplitude of
the $p$-th beam is $E_p$. $q=0,\pm 1$ correspond to $\pi$, $\sigma^\pm$ polarizations components respectively in each laser beam.
The polarization vectors under the Cartesian coordinate axis are: $\hat{\epsilon}_0=\hat{e}_z$, $\hat{\epsilon}_\pm=\mp(\hat{e}_x\pm i\hat{e}_y)/\sqrt{2}$.
$M_{pq}$ gives the fraction for each polarization component in the $p$-th beam. 

Using Eq.(\ref{eq1}), the Hamiltonian under interaction picture reads as
\begin{equation}\label{eq2}
 H = \frac{\hbar}{2}\sum\limits_{i,j}\sum_{q} \Omega_{ij}^{(q)}|j\rangle\langle i| + \text{h.c.},
\end{equation}
with $|i\rangle$ and $|j\rangle$ indicate sublevels in the ground and excited states respectively, and 
\begin{equation}\label{eq3}
 \Omega_{ij}^{(q)} = \sum\limits_p e^{i\vec{k}_p\cdot\vec{r}} 
 e^{-i\Delta_{pq}^{ij}(t)t}M_{pq}A_{ij}^{(q)}\Omega_p.
\end{equation}
Here $\Omega_p$ is the total on-resonance Rabi frequency of the $p$-th beam. 
The detuning $\Delta_{pq}^{ij}(t) = \omega_p(t) - \omega_{ij} + \delta_{ij}(B)$ with 
$\omega_p(t)$ in a sawtooth shape shown in 
Fig.\ref{fig1}(b), and $\omega_{ij}$ is the resonant frequency 
for the $|i\rangle \to |j\rangle$ transition, $\delta_{ij}(B)=\delta_i(B)-\delta_j(B)$ where $\delta_{i(j)}(B)$ is the
energy shift of the sublevel $|i\rangle(|j\rangle)$ under a magnetic field strength of $B$ [see Fig.\ref{fig1}(d)]. 
$A_{ij}^{(q)}$ are the matrix elements for all electric dipole allowed transitions. 
For BaF molecule, under a weak magnetic field, $B<10~\text{G}$, $F$ is a good quantum number, 
the derivations in Ref.\cite{Chen2016} can be directly applied. However, in strong magnetic field regime, for example, 
$B=100~\text{G}$, the selection rules change.
We label the sublevels in ground $X$ state in fully decoupled basis $|S,m_S;N,m_N; I, m_I\rangle$ while
for those in excited state we use $|J,m_J;I,m_I\rangle$. The derivations and values of $A$ are summarized in appendix \ref{appA}.

The force from the interactions with the light field can be yeilded as 
a gradient of the total energy of the system, that is,
\begin{equation}
 \hat{F} = -\nabla H = -\hbar\sum\limits_{i,j}\sum_q(\nabla\Omega_{ij}^{(q)})|j\rangle\langle i| + \text{h.c.} ,
\end{equation}
then the expectation value of the force for a system represented 
by a density matrix $\rho$ is $\langle F\rangle = 
\text{Tr}(\rho \hat{F})$. The time evolution of the density matrix 
$\rho$ is determined by the master equation
\begin{equation}
 \frac{\partial \rho}{\partial t} = \frac{1}{i\hbar}[H(t),\rho] 
 + \frac{1}{2}\sum\limits_{ij}(2C_{ij}\rho C_{ij}^\dagger 
 - C_{ij}^\dagger C_{ij}\rho - \rho C_{ij}^\dagger C_{ij}),
\end{equation}
with the collapse operator $C_{ij} = \sum_q A_{ij}^q\sqrt{\Gamma}|i\rangle\langle j|$. 
We calculate the forces in one period when the density matrix reaches quasi-steady. 

For laser cooling with the type-II transitions and molecules, an
angled magnetic field $B$ should be introduced to elimilate the dark states. 
The direction of the magnetic field redefines the local quantum axis,
and we assume the field on the coordinate $yz$ plane with an angle
of $\theta$ to the $z$ axis.  The new fraction vector for the three polarizations
in each laser beam can be transformed from the old one, i.e.,
\begin{equation}\label{eq6}
\left[\begin{array}{l} M'_{+1} \\ M'_0 \\
M'_{-1}\end{array}\right] = \frac{1}{2}\left[\begin{array}{ccc}1+\cos\theta & \sqrt{2}\sin\theta & 1-\cos\theta \\
-\sqrt{2}\sin\theta & 2\cos\theta & \sqrt{2}\sin\theta \\
1-\cos\theta & -\sqrt{2}\sin\theta & 1+\cos\theta \end{array}\right]\left[\begin{array}{l}M_{+1} \\
M_0 \\ M_{-1}\end{array}\right].
\end{equation}

In our calculations below, we consider a light field from two counter-propagating beams in $\sigma^+-\sigma^-$ configuration, and
the two beams have equal Rabi frequencies $\Omega$. 
Under an angled magnetic field ($\theta\neq 0$),
it contains all the three polarization components and 
thus can effectively eliminate the dark states. We first investigate the possible type-II
transitions in Fig.\ref{fig1}(c) under weak magnetic field, and then turn our attention to strong magnetic field regime, finally to the BaF molecule.
For BaF molecule, the linewidth of the $A^2\Pi_{1/2}$ excited state is 
$2\pi\times 2.84~\text{MHz}$, and the hyperfine splittings under magnetic field are shown in Fig.\ref{fig1}(d). 
We will try to analyze the dependences of the SWAP cooling on the sweeping parameters, including the repetition
rate and the frequency ramp speed, and perform a comparison with the conventional Doppler cooling scheme.

\section{SWAP cooling for type-II transitions}\label{sec3}

\subsection{Weak magnetic field regime}\label{sec3A}

\begin{figure}[]
\includegraphics[width=0.48\textwidth]{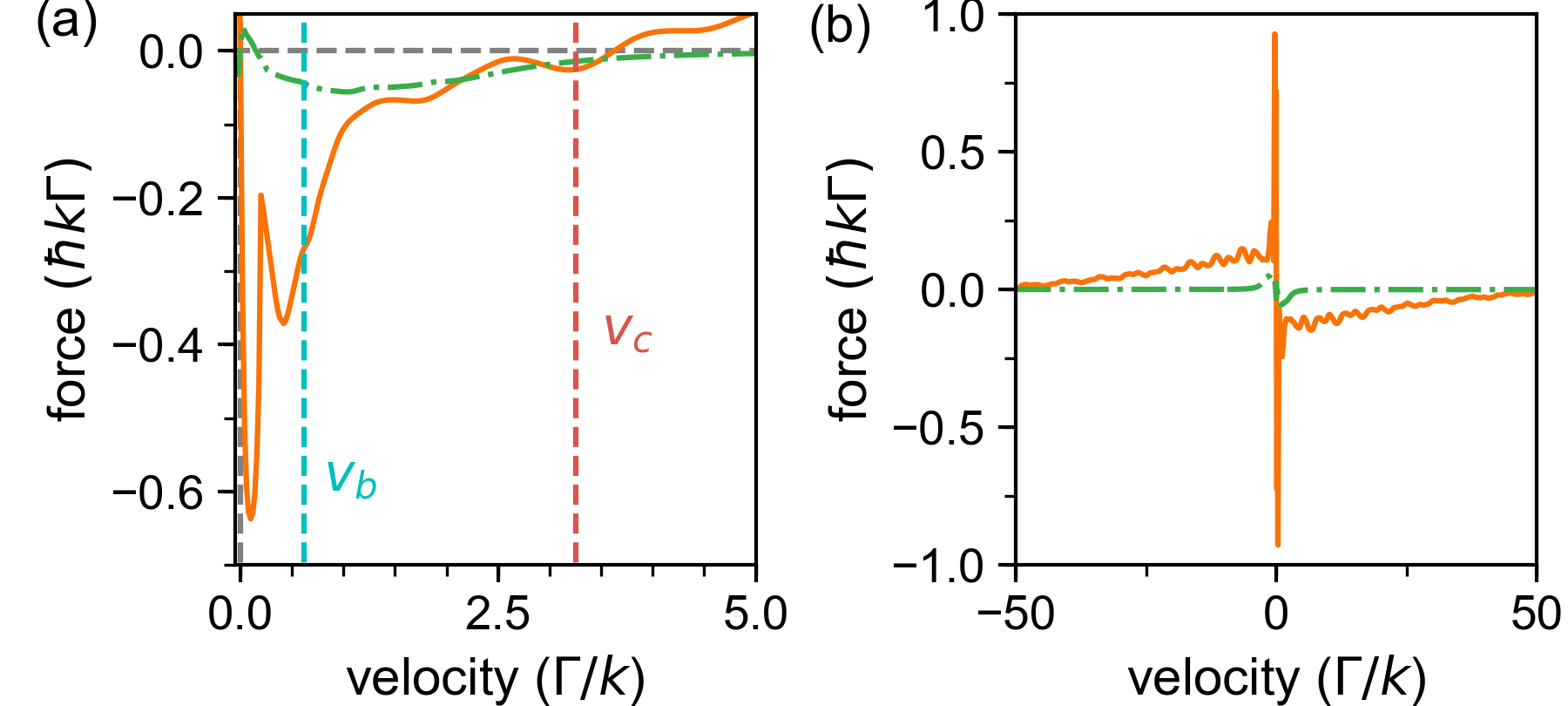}
\caption{\label{fig2} (Color online) (a) A comparison of the force from the SWAP scheme (orange solid line) and that from the
conventional cooling (green dotted dashed line). The SWAP parameters are: $\Omega=10\Gamma$, $\alpha=2\Gamma^2$, $T=5\Gamma^{-1}$ and 
$B=5~\text{G}~(\theta=\pi/2)$. For conventional Doppler cooling, $\Omega=2\Gamma$, $\Delta=-\Gamma$ and 
$B=5~\text{G}~(\theta=\pi/2)$. The critical value of $v_c$ indicated by the red dashed line
is obtained from Eq.(\ref{eq7}), and the cyan dashed line indicates the critical velocity $v_b$ determined from Eq.(\ref{eq8}). 
(b) The SWAP force (orange solid line) with a large frequency ramp speed $\alpha=50\Gamma^2$. The other parameters are: $\Omega=50\Gamma$,
$T=5\Gamma^{-1}$ and $B=5~\text{G}(\theta=\pi/2)$. The conventional Doppler cooling force with the same parameters in (a) is shown in green dotted
dashed line.}
\end{figure}

In the quasi-closed cycling transition for the diatomic molecules under a weak magnetic field, as shown in Fig.\ref{fig1}(c), there are
three different multi-level type-II transitions: (i) $F=1\to F'=0$, (ii) $F=1\to F'=1$ and (iii) $F=2\to F'=1$. We first focus on the 
case (i) and use the parameters for the $F=1^{(-)}\to F'=0$ transitions in the BaF molecule. The Land\'e $g$-factor for the ground state is $g_F = -0.51$, and a 
magnetic field of $B=5~G$ ($\theta=\pi/2$) leads to a Zeeman splitting of $\Delta_z=1.25\Gamma$. 

To check the effect of the magnetic field on the SWAP cooling, we first consider a low laser intensity case, i.e., a small
Rabi frequency $\Omega=10\Gamma$. To fulfill the adiabatic condition $|A_{ij}^{(q)}\Omega|^2\gg\alpha$, we choose 
the frequency ramp speed $\alpha=2\Gamma^2$. The sweeping period $T=5\Gamma^{-1}$. 
Figure \ref{fig2}(a) shows a comparison between the SWAP force and  the conventional Doppler damping force. They show different features. First, the coolable velocity region of the SWAP scheme 
has an upper limit $v_c$. The SWAP cooling requires that the frequency sweeping covers the resonant positions for all possible transitions \cite{Bartolotta2018},
this critical limit $v_c$ is determined by both the sweeping range $\Delta_\text{T}=\alpha T$ and the Zeeman shift $\Delta_z$, that is, 
\begin{equation}\label{eq7}
v_c = \frac{\Delta_\text{T}-2\Delta_z}{2k}.
\end{equation}
Second, an enhancement of the cooling force appears in the small velocity region. For conventional Doppler cooling with typical parameters, 
as shown in Fig.\ref{fig2}(a), the maximum achievable cooling force approximates $\sim0.06\hbar k\Gamma$. However, this value under the SWAP 
scheme is about 10 times larger, $\sim 0.6\hbar k\Gamma$. It happens for small velocities less than a critical value $v_b$; see Fig.\ref{fig2}(a).
Such a strong cooling ability should resort to the nontrivial Bragg oscillations that can induce more than $2\hbar k$ of momentum transfer 
during one sweeping period for a two-level system, as discussed in Ref.\cite{Bartolotta2018}. Distinct from the two-level case, in a multi-level system, the
enhanced cooling region is resitricted by the energy splittings between each pair of sublevels.

\begin{figure}[b]
\includegraphics[width=0.45\textwidth]{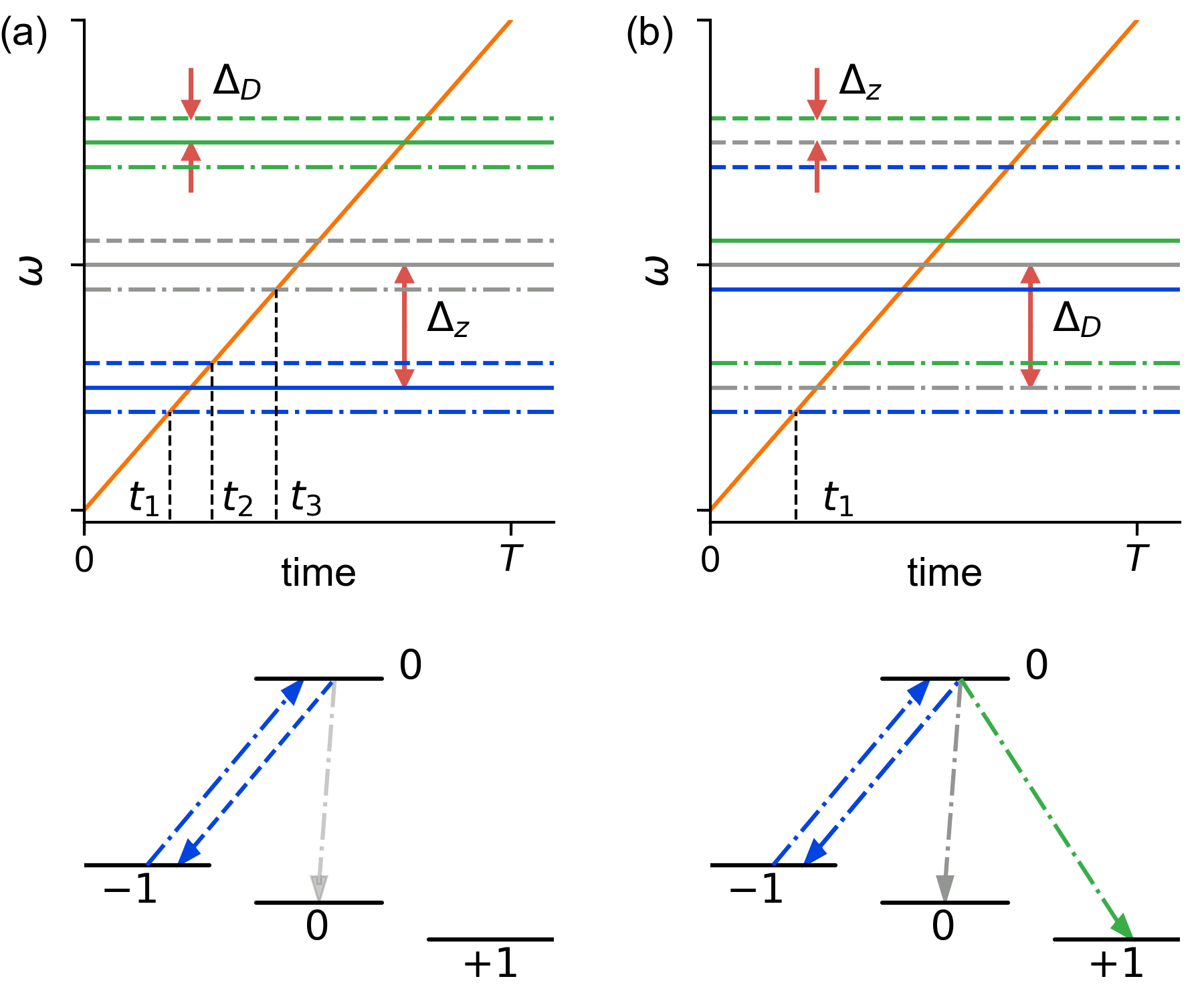}
\caption{\label{fig3}(Color online) Resonant positions for each polarization component in the two beams for the case of: (a) $\Delta_z\gg\Delta_\text{D}$ and 
(b) $\Delta_\text{D}\gg\Delta_z$. The solid lines indicate the resonant frequencies of the three transitions without considering the Doppler effect. The three
polarizations are depicted in different colors: $\sigma^+$ in blue, $\pi$ in gray, and $\sigma^-$in green. The dashed lines (dotted dashed lines) indicate the resonant positions of the
three polarization components in the co-propagating (counter-propagating) beam, all shift $+\Delta_\text{D}$ ($-\Delta_\text{D}$) compared to the solid lines. $t_{1}$ ($t_2$) is the
time when the $\sigma^+$ component in the counter-propagating (co-propagating) beam becomes resonant with the $m_F=-1\to m'_F=0$ transition. 
In (a), $t_3$ is the time when the $\pi$ component in the counter-propagating beam becomes resonant with the $m_F=0\to m'_F=0$ transition. 
The excitation and deexcitation processes after $t_1$are schematically plotted below respectively. The linestyles have the same meanings with those in upper panels.}
\end{figure}

To determine $v_b$, let us analyze the different cooling mechanisms for large and small velocities in the SWAP scheme for the
$F=1\to F'=0$ transition. For a small velocity, the Doppler shift $\Delta_\text{D}=kv$ is smaller than the Zeeman shift $\Delta_z$, as shown in Fig.\ref{fig3}(a),
the time sequence of the stimulated excitation and deexcitation of the three transitions is still in order. At time $t_1$ in Fig.\ref{fig3}(a), the particles in
Zeeman sublevel $m_F=-1$ get excited by the $\sigma^+$ component in the counter-propagating beam, following which the stimulated deexcitation 
back to $m_F=-1$ happens as the frequency ramps to be resonant with the $\sigma^+$ polarization component in the co-propagating beam at time $t_2$. This is similar to the two-level case.
However, when $t_3$ is close to $t_2$ [see Fig.\ref{fig3}(a)], the deexcitation to the $m_F=0$ sublevel simultaneously happens with photons emitted in the same direction of the moving velocity,
leading to a cancellation of the momentum exchange from the stimulated excitation at time $t_1$. Competition between the deexcitation to $m_F=-1$ via the co-propagating beam and the deexcitation to $m_F=0$
via the counter-propagating beam certainly suppresses the net momentum transfer in one sweeping period. The critical velocity $v_b$ is defined when $t_2=t_3$. Taking both the Doppler shift and the Zeeman shift into consideration, we have $\omega_0-\Delta_z+\Delta_\text{D}=\omega_0-\Delta_\text{D}$, resulting in
\begin{equation}\label{eq8}
v_b = \Delta_z/2k.
\end{equation}  
In Fig.\ref{fig2}(a), a rapid decrease of the SWAP cooling force for velocities larger than $v_b$ is consistent with the above discussion. 

To depict a clear picture of the momentum transfer 
for large velocities where $\Delta_\text{D}\gg\Delta_z$, we perform calculations with a large frequency ramp speed $\alpha=50\Gamma^2$, and accordingly a large Rabi frequency
of $\Omega=50\Gamma$ to fulfill the adiabatic condition. The results are shown in Fig.\ref{fig2}(b). The cooling region indeed become wider up to
$50\Gamma/k$, consistent with Eq.(\ref{eq7}). The maximum cooling force approximates $\sim \hbar k\Gamma$ due to strong Bragg oscillation effect with a large $\Omega$ \cite{Bartolotta2018}. However, 
for velocities larger than $5\Gamma/k$, the cooling forces are all below $0.2\hbar k\Gamma$, which means that the net momentum transfer is 
less than $\hbar k$ during one sweeping period of $T= 5\Gamma^{-1}$. This can be easily understood from Fig.\ref{fig3}(b). At time $t_1$, the $\sigma^+$ component in the counter-propagating beam
becomes on resonance with the $m_F=-1\to m'_F=0$ transition. Since the Doppler shift is large, before the co-propagating beam reaches on resonance, the deexcitations back to the three ground sublevels
already happen as the $\pi$ and $\sigma^-$ components in the counter-propagating beam first become near resonance and 
drive the stimulated emission processes from $m'_F=0$ to $m_F=0$ and $m_F=+1$ respectively. Therefore, the exchanged
momenta from excitation and deexcitation are in opposite directions, making the net momentum transfer smaller than $\hbar k$. For a finite velocity $v$, to make the deexcitation from co-propagating beam
play a role, we roughly give a reasonable condition: the Rabi frequency $|A_{ij}^{(q)}\Omega|>kv$, which might provide a direction for slowing a multi-level particle beam with the SWAP scheme. 

From the above discussions, we conclude that, in a multi-level transition, an excitation by the counter-propagating beam must be strictly
followed by an emission stimulated by the co-propagating beam to ensure a considerable momentum transfer. Anything that contributes to mix the beam roles or simply leads to absorb and reemit photons 
within a single beam will necessarily degrade the quality of the cooling process. The angled magnetic field remixes the dark states and effectively widen the enhanced cooling region, as the 
energy splittings from the magnetic field make the three transitions well isolated with each other. Ideally, by assuming that the three transtions are excited and deexcited independently and considering the
branching ratios, the net momentum transfer during one sweeping period should be $2\hbar k$, leading to the maximum force limit of $F_\text{max}=2\hbar k/T$. However, under weak magnetic field, this
limit is masked by the Bragg oscillation effect since both occur in the small veloticy region.

Another issue that should be kept in mind is the role of the spontaneous emission in the sweeping process. A low probability of spontaneous decay is required to realize an
effective SWAP cooling, that is, the time $t_e$ that the particle spends in the excited state should be much smaller than the lifetime $\Gamma^{-1}$. Similar to the two-level
system \cite{Bartolotta2018}, the time $t_e$ can be estimated with the time interval between the two resonant time points of the counter-propagating and co-propagating beams, 
i.e., $t_e\sim 2kv/\alpha$. Then, we have another condition, $v\ll \alpha\Gamma^{-1}/2k$. This should be taken into consideration in simulations for diatomic molecules,
especially those with leakage channels, where the spontaneous emission should be suppressed as effectively as possible.

\begin{figure}
\includegraphics[width=0.48\textwidth]{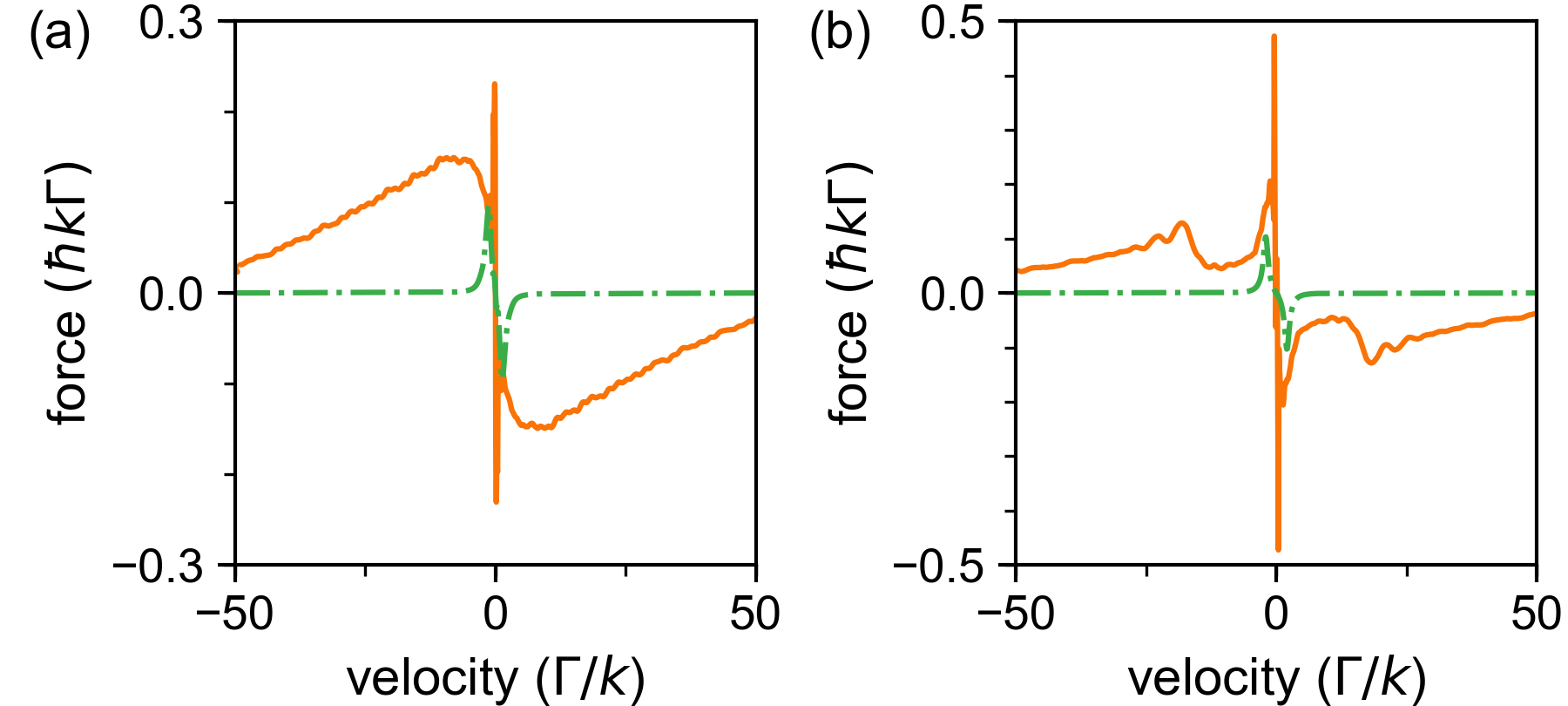}
\caption{\label{fig4} (Color online) The velocity dependent SWAP forces (orange solid lines) for two multi-level type-II transitions: (a) $F=1\to F'=1$ and 
(b) $F=2\to F'=1$. The SWAP parameters are: $\Omega=50\Gamma$, $\alpha=50\Gamma^2$ and $T=5\Gamma^{-1}$. The conventional Doppler cooling forces (green dotted dashed lines) are 
also shown respectively with $\Omega=2\Gamma$, $\Delta=-\Gamma$. For both the SWAP and the Doppler cooling schemes,
the angled magnetic field is $B=5~\text{G}~(\theta=\pi/2)$. The Land\'e $g$ factors used are: $g_{F=1}=-0.51$, $g_{F=2}=1.01$ and $g_{F'=1} = -0.2$.}
\end{figure}

Besides the $F=1\to F'=0$ transition, the SWAP scheme under weak magnetic field on other two type-II transitions in Fig.\ref{fig1}(c), $F=1\to F'=1$ and $F=2\to F'=1$, has also been investigated, as
shown in Fig.\ref{fig4}. The maximum achievable cooling forces for both two transitions are smaller than $0.5\hbar k\Gamma$. The enhancements are not as significant as that in the $F=1\to F'=0$ 
transition [see Fig.\ref{fig2}(b)] due to more Zeeman sublevels that make the in-order excitation and deexcitation much more fragile as 
more transitions are involved. However, for large velocities, the cooling forces are still below $0.2\hbar k\Gamma$, and the cooling mechanism is similar to that discussed in the $F=1\to F'=0$ transition.
Together with the experimetal demonstration of the SWAP cooling scheme on the type-I $F=0\to F'=1$ transtion \cite{Muniz2018}, the results here indicate that the SWAP cooling might be applicable to molecules
under a weak magnetic field. We will discuss this in Sec.\ref{sec4}. 

\subsection{Strong magnetic field regime}

Inspired by the suggestion in Sec.\ref{sec3A} that larger energy gaps between each two sublevels lead to a wider enhanced cooling velocity range, we investigate the
SWAP force on possible transitions related to the molecules under a strong magnetic field. If the total angular momentum $F$ is still a good quantum number
under a large magnetic field of $100~\text{G}$, then the enhanced cooling velocity limit would be large and still determined by Eq.(\ref{eq8}), 
about $\sim 30~\text{m/s}$ with 
the parameters for BaF. However, according to Fig.\ref{fig1}(d), the energy shift of each sublevel no longer linearly varies and $F$ is not a good quantum
number yet. The selection rules also change, and the calculation on the matrix elements of the electric dipole transitions tells us that the 4+12 level structure can be
decomposed into four 1+5 subsystems; see Appendix \ref{appA}. Therefore, the discussions on the momentum transfer in the weak magnetic field regime becomes invalid. 

\begin{figure}[]
\includegraphics[width=0.48\textwidth]{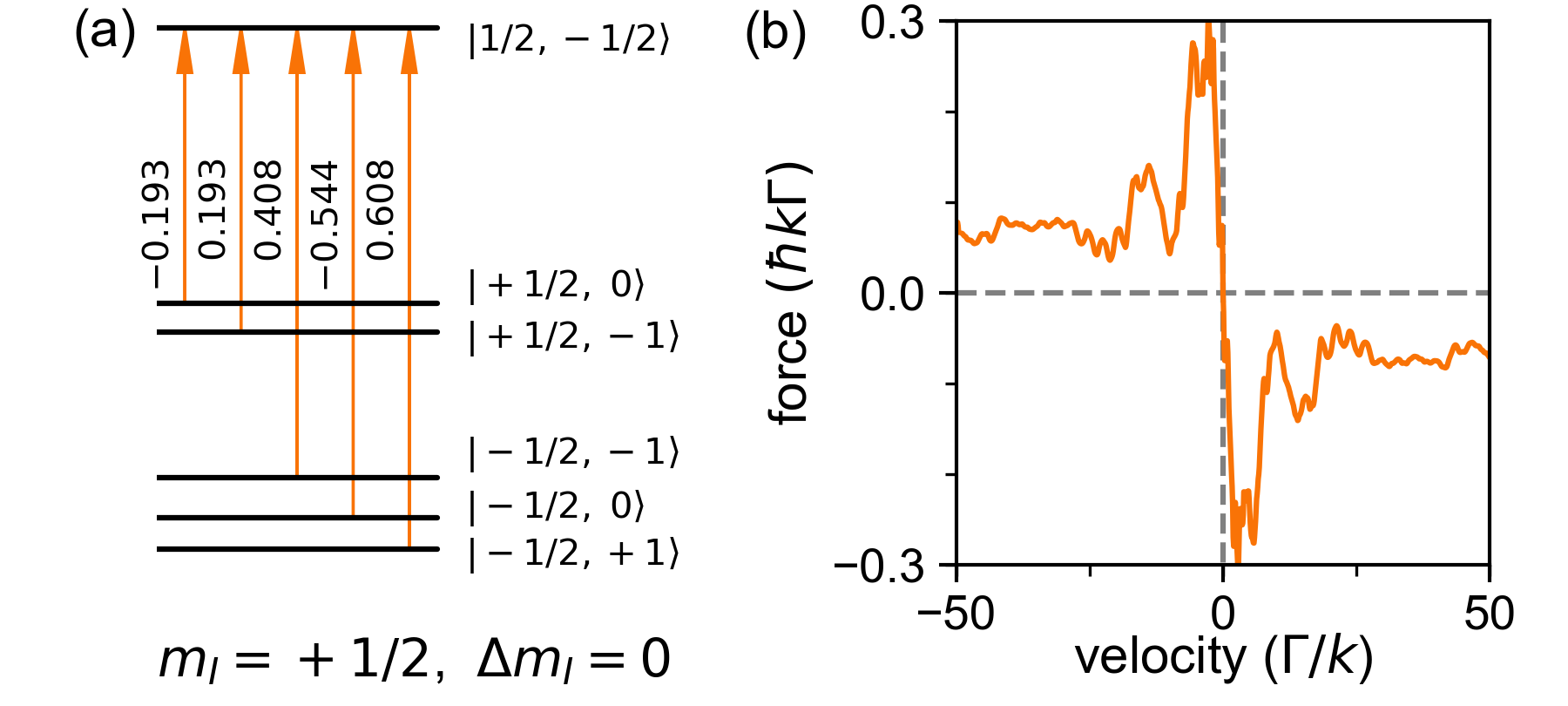}
\caption{\label{fig5}(Color online) (a) The 1+5 system with the excited state $|J=1/2, m_J=-1/2; I=1/2, m_I = +1/2\rangle$ in the BaF molecule. All five possible
transitions from the ground state labelled by $|m_S, m_N\rangle$ are shown. The transition strengths, i.e., the matrix elements of $A$, are from Table \ref{tabA}.
(b) The SWAP cooling force under a magnetic field of $B=100~\text{G}~(\theta=\pi/2)$. The SWAP parameters are: $\Omega=100\Gamma$, $\alpha
=100\Gamma^2$, and $T=3\Gamma^{-1}$. The detuning of the center frequency $\Delta=0$, i.e., the center frequency is the gap between the zero energy point in
Fig.\ref{fig1}(d) and the excited state.
}
\end{figure}

We take the 1+5 subsystem with the excited state $|J=1/2, m_J=-1/2; I=1/2, m_I = +1/2\rangle$ as an example. Figure \ref{fig5}(a) shows all five possible 
electric dipole allowed transitions from the ground state labelled by $|m_S, m_N\rangle$ with $m_I=+1/2$ since the selection rules require $\Delta m_I=0$. 
The energy gap between the highest and the lowest sublevels in the ground state is $\sim 130\Gamma$ ($B=100~\text{G}$) leads to a large frequency sweeping range $\Delta_\text{T}$. 
Here we choose the sweeping speed $\alpha=100\Gamma^2$ and the sweeping period 
$T=3\Gamma^{-1}$. To fulfill the adiabatic condition $|A_{ij}^{(q)}\Omega|^2\gg\alpha$, the total Rabi frequency $\Omega=100\Gamma$ is used. 

The calculated SWAP force is shown in Fig.\ref{fig5}(b). Behaviours similar to the weak magnetic field case are depicted, i.e., a 
considerable enhancement of the cooling force in the small velocity region and a large coolable velocity range. The enhanced cooling velocity limit $v_b$ still
depends on the energy gaps in the ground state. 
When $B=100~\text{G}$, the energy gap between $|m_S=-1/2, m_N=-1\rangle$ and 
$|m_S=+1/2, m_N=-1\rangle$ sublevels is $\sim 80\Gamma$ and other gaps between every two neighbouring sublevels are about $\sim 15\Gamma$. 
Then, $v_b\sim 7.5\Gamma/k$ is estimated from Eq.(\ref{eq8}), in good agreement with the result shown in Fig.\ref{fig5}(b). For velocities larger than $v_b$, the excitations by 
the counter-propagating beam and deexcitations from the co-propagating beam become out of order. The second dip around $\sim 20\Gamma/k$ in Fig.\ref{fig5}(b)
might be induced by the different transition strengths (see the values of $A_{ij}^{(q)}$), since the net momentum transfer depends on which deexcitation, 
from the co-propagating beam or the counter-propagating beam, dominates during one sweeping period. 

From Table \ref{tabA} in Appendix \ref{appA}, the other three 1+5 subsystems have similar transition strengths to those in Fig.\ref{fig5}(a). 
Therefore, we roughly expect similar features of the SWAP cooling forces for these 1+5 subsystems, which indicates that SWAP cooling 
molecules under a large magnetic field is possible.

\section{SWAP cooling for molecules}\label{sec4}

\begin{figure}[b]
\includegraphics[width=0.48\textwidth]{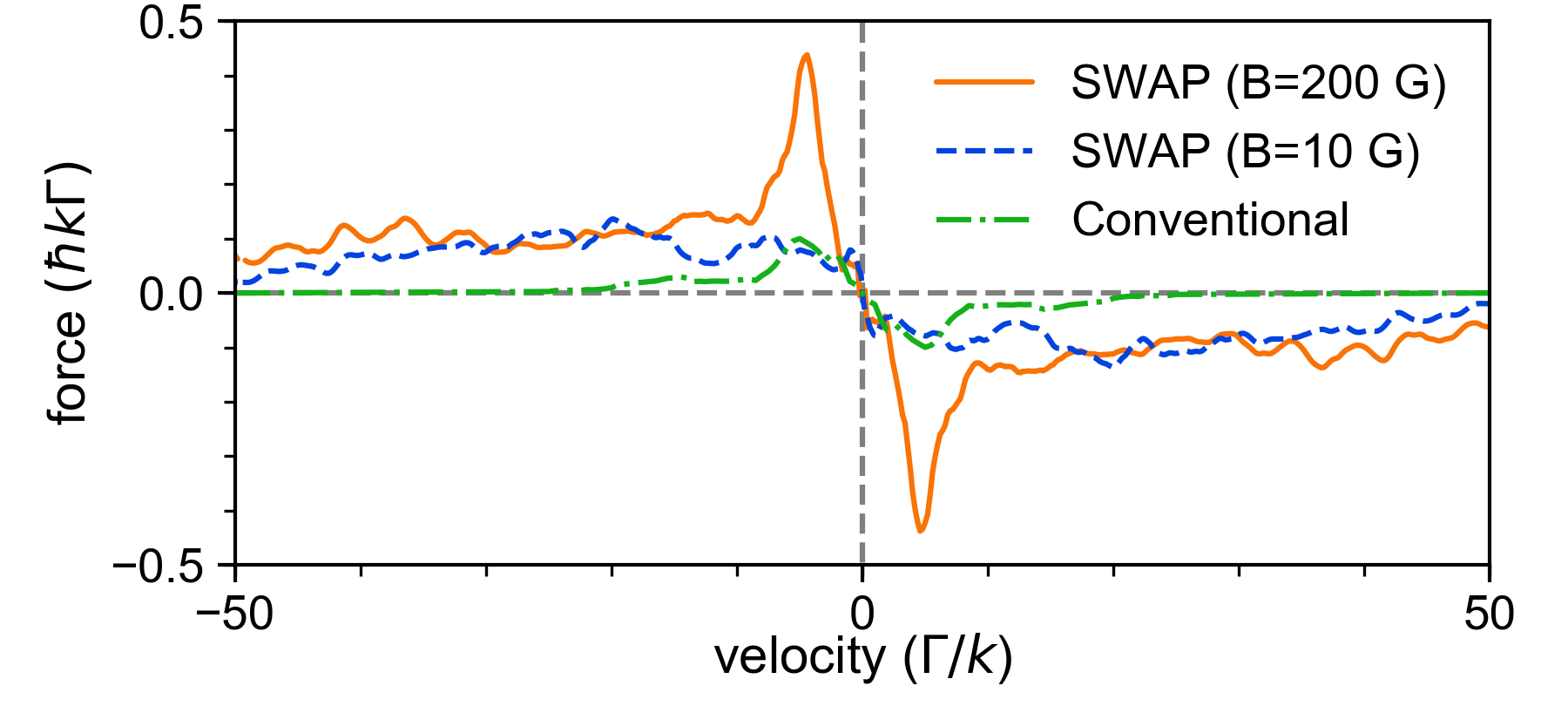}
\caption{\label{fig6}(Color online) Comparisons of the SWAP forces to the conventional Doppler cooling force. 
Under a weak magnetic field of $B=10~\text{G} (\theta=\pi/2)$ (blue dashed line), the SWAP parameters are: $\Omega=100\Gamma$, $\alpha=50\Gamma^2$, 
$T=3\Gamma^{-1}$, $\Delta = +7\Gamma$. With a strong magnetic field of $B=200~\text{G}$ (orange solid line), the SWAP parameters are: 
$\Omega=50\Gamma$, $\alpha=80\Gamma^2$, $T=4\Gamma^{-1}$, $\Delta = -25\Gamma$. The parameters used for conventional Doppler cooling
(green dotted dashed line) are: the Rabi frequency $\Omega=5\Gamma$, the detuning $\Delta=-5\Gamma$, the angled magnetic field $B=5~\text{G} (\theta = \pi/2)$, and
a 38 MHz sideband modulation is also applied to cover all hyperfine states for the BaF molecule. 
}
\end{figure}

The results in the multi-level type-II transitions in Sec.\ref{sec3} indicate that application of the SWAP scheme on molecules might be possible. One problem
in a diatomic molecule arises from the large hyperfine splitting. For example, with no external magnetic field applied, the energy gap $\Delta_\text{hf}$ 
between the $F=1^{(-)}$ state and the $F=2$ state is about $\sim 56\Gamma$ for the BaF molecule \cite{Chen2016}. To cover all sublevels, a large frequency
sweeping range $\Delta_\text{T}=\alpha T$, and therefore a large total Rabi frequency $\Omega$ are required. Again, we first consider the case with a 
weak magnetic field $B=10~\text{G}$. As shown in Fig.\ref{fig6}, even with $\Omega=100\Gamma$, the maximum SWAP force is about $0.1\hbar k\Gamma$, at the
same magnitude with that for the conventional Doppler cooling. Nevertheless the enhancement does not 
appear, a wider coolable veloticy region is still observed and the critical velocity $v_c$ is determined by the sweeping range $\Delta_\text{T}$.
For the parameters used in Fig.\ref{fig6}, $v_c\approx (\Delta_\text{T}-\Delta_\text{hf})/2k \sim 50\Gamma/k$ with $\Delta_\text{T}=150\Gamma$.

The behaviours of the SWAP force under a large magnetic field are different; see Fig.\ref{fig6}. 
Compared to the conventional Doppler scheme, the maximum achievable cooling force
is $4\times$ larger, that is, $\sim 0.4\hbar k\Gamma$. 
The enhanced cooling velocity limit $v_b$ which is determined by the
energy gaps between every two neighbouring sublevels, as discussed in Sec.\ref{sec3}. From Fig.\ref{fig1}(d), the energy shift for each sublevel 
under larger magnetic field approximately varies in parallel with each other in the same $m_S$ branch. When the magnetic field
strengh becomes larger than $100~\text{G}$, the energy gaps 
maintain around $\sim 15\Gamma$, leading to $v_b\sim7.5\Gamma/k$, which is consistent with the result in Fig.\ref{fig6}.
On the other hand, the SWAP force has a weak velocity selective character, and the cooling forces for large velocities are similar to those 
with $B=10~\text{G}$, approximate $\sim 0.1\hbar k \Gamma$. The coolable velocity region below $v_c$ is determined by both $\Delta_\text{T}$ and
$B$. The critical value of $v_c$ is obtained from
\begin{equation}\label{eq9}
2kv_c + \Delta_\text{hf} + \Delta_s(B) = \Delta_\text{T},
\end{equation}
where $\Delta_s(B) = \mu_B g_s B/\hbar$ with $g_s\approx2$ is the energy splitting between the two $m_S=\pm 1/2$ branches with the magnetic 
field strength in unit of G. 

\begin{figure}[b]
\includegraphics[width=0.48\textwidth]{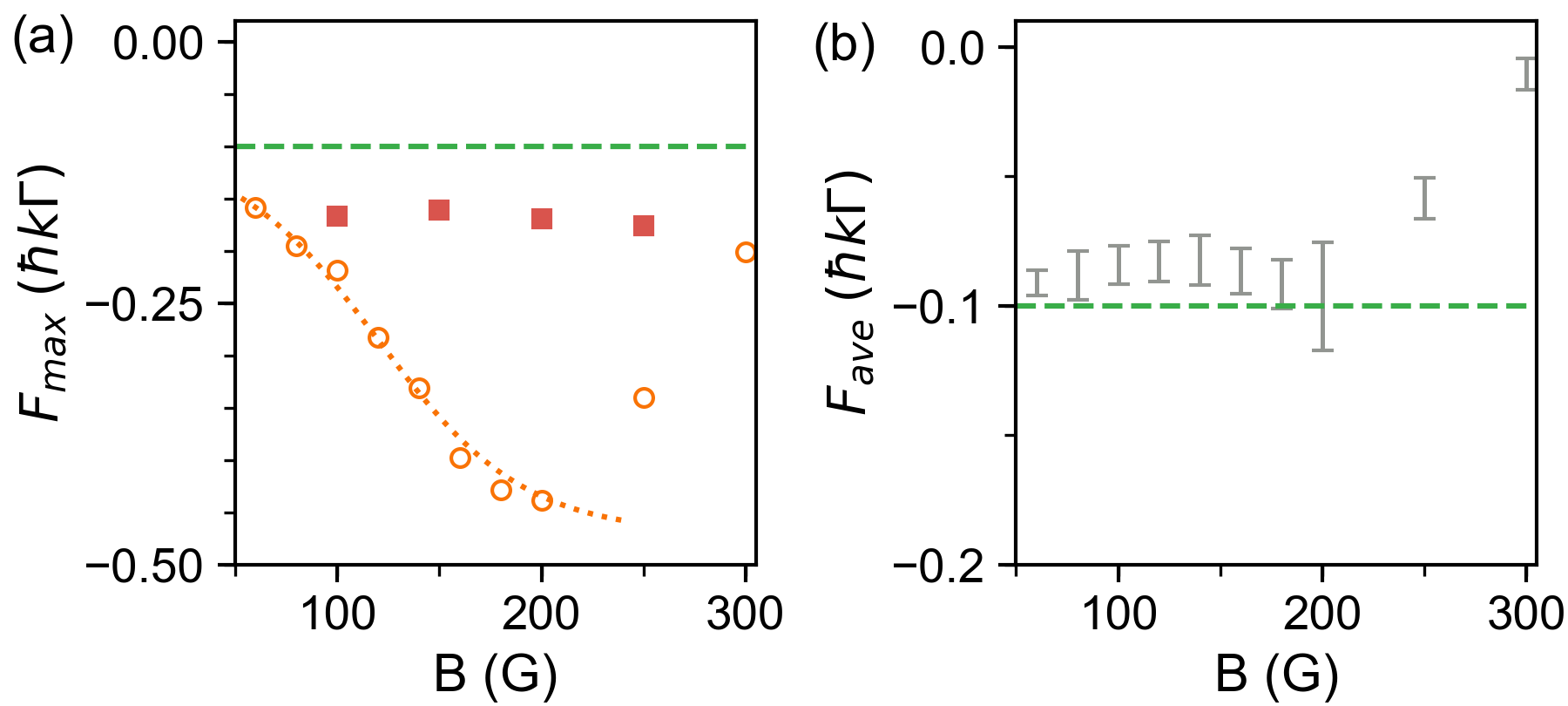}
\caption{\label{fig7}(Color online) (a) The maximum achievable cooling force $F_\text{max}$ under different magnetic field strength. The paramters used for the 
SWAP scheme (orange circle) is: $\Omega=50\Gamma$, $\alpha=80\Gamma^2$, $T=4\Gamma^{-1}$, $\Delta=-28\Gamma$. The red square points are calculated under the same 
SWAP scheme but with assumed degenerate sublevels in the excited state. The dotted line is a guide to eyes. 
(b) The average force $F_\text{ave}$ for large velocities from $30\Gamma/k$ to $50\Gamma/k$ under 
different magnetic field strength. The green dashed lines in both (a) and (b) indicate the value of $F_\text{max}$ in the conventional Doppler cooling from 
Fig.\ref{fig6}.
}
\end{figure}

\begin{figure*}[]
\includegraphics[width=0.95\textwidth]{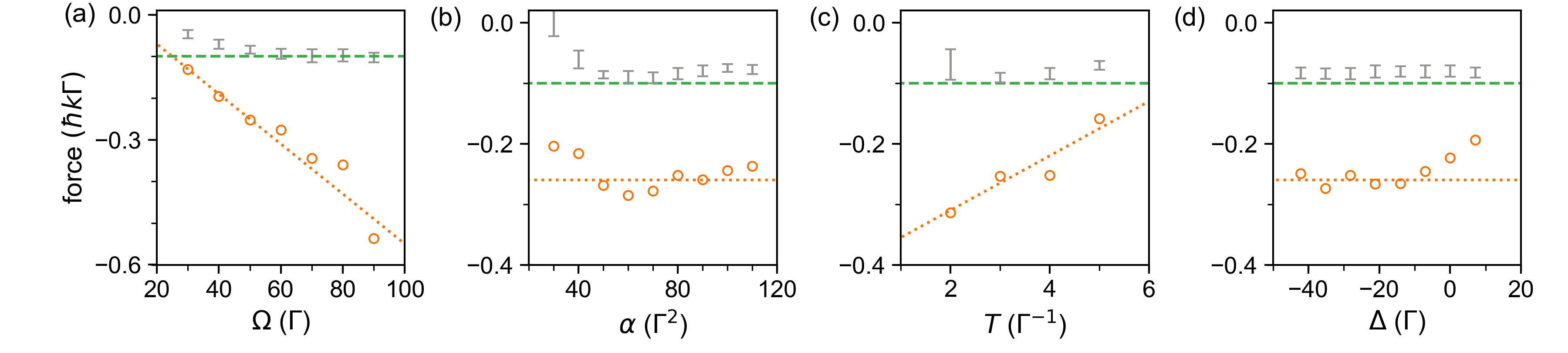}
\caption{\label{fig8}(Color online) Dependences of the maximum achievable cooling force $F_\text{max}$ (orange circle) and the average force $F_\text{ave}$ 
(gray point) for large velocities from $30\Gamma/k$ to $50\Gamma/k$ on the parameters of the frequency chirped cooling lasers: (a) the total Rabi frequency $\Omega$,
(b) the frequency ramp speed $\alpha$, (c) the sweeping period $T$ and (d) the shift $\Delta$ of the center frequency $\omega_0$. The green dashed lines indicate
the value of the maximum achievable cooling force under the Doppler cooling scheme with the same parameters in Fig.\ref{fig6}. The orange dotted lines are guides to eyes. 
The magentic field strength in the SWAP scheme is $B=100~\text{G}$. The SWAP parameters used are: 
(a) $\alpha=80\Gamma^2$, $T=4\Gamma^{-1}$, $\Delta=-28\Gamma$, (b) $\Omega=50\Gamma$, $T=4\Gamma^{-1}$, $\Delta=-28\Gamma$, (c) $\Omega=50\Gamma$, $\alpha=80\Gamma^2$, 
$\Delta=-28\Gamma$, (d) $\Omega=50\Gamma$, $\alpha=80\Gamma^2$, $T=4\Gamma^{-1}$.
}
\end{figure*}

To systematically study the SWAP force with different parameters, we employ two variables to describe the cooling ability. One is the maximum achievable cooling force 
$F_\text{max}$, and another is the average force $F_\text{ave}$ for large velocities from $30\Gamma/k$ to $50\Gamma/k$. As illustrated in Fig.\ref{fig7}(a),
by increasing the magnetic field strength from $60~\text{G}$ to $200~\text{G}$, although the enhanced cooling region does not change a lot [due to 
little change of the value $v_b$], the $F_\text{max}$ increases significantly from $0.15\hbar k\Gamma$ to $0.4\hbar k\Gamma$. We attribute such a phenonmenon
to the energy splitting between sublevels with different $m_J$ in the excited state. To prove this assertion, we perform 
another series of calculations with assumed degenerate sublevels in the excited state, and the resulting $F_\text{max} \sim 0.17\hbar k\Gamma$
is smaller and insensitive to the value of $B$; see Fig.\ref{fig7}(a). Here we roughly give a lower limit of the magnetic field strength when the energy splitting of the excited state is at the same 
magnitude with that in the ground state [labelled with $\Delta_g$], i.e., $\mu_B g_e B/\hbar > \Delta_g$. 
For BaF molecule, $\Delta_g\sim 15\Gamma$, $g_e=-0.2$, we have
$B> 150~\text{G}$. This is consistent with our calculations in Fig.\ref{fig7}(a) as the $F_\text{max}$ slowly increases after $B$ reaches this value. 

However, the magnetic field is also restricted by the frequency sweeping range. For $B>200~\text{G}$, a sharp decreasement of the $F_\text{max}$ appears in
Fig.\ref{fig7}(a) since the frequency sweeping range can not cover all possible transitions any more. Therefore, the upper limit of the magnetic field is 
determined from Eq.(\ref{eq9}). This can also be easily figured out from Fig.\ref{fig7}(b) where the average force $F_\text{ave}$ for large velocities approaches zero
for $B>200~\text{G}$. As a result, we conclude that a suitable magnetic field is required to achieve both a considerable enhancement of the maximum cooling force and
a large coolable velocity region.

Figure \ref{fig8} shows the dependences of the $F_\text{max}$ and $F_\text{ave}$ on the SWAP parameters. For a fixed frequency ramp speed $\alpha$, increasing
the Rabi frequency $\Omega$ within the adiabatic condition $\Omega^2\gg\alpha$ results in approximately linear increasing of the $F_\text{max}$; see 
Fig.\ref{fig8}(a). It is clear that the $F_\text{max}$ becomes two times larger than that from the conventional Doppler cooling scheme when $\Omega^2>20\alpha$. 
Once this critical condition is fulfilled, the cooling forces for large velocities also approach the maximum value of the Doppler cooling force, but $F_\text{ave}$
maintains nearly a constant even with a rather large $\Omega$. 
Increasing the Rabi frequency $\Omega$ only affects the cooling forces in the velocity region below $v_b=\Delta_g/2k$. It is not likely
to achieve an enhanced cooling effect for large velocities by roughly increasing the laser intensity, which should be kept in mind if one expects using the SWAP 
force to slow a molecular beam.

In Fig.\ref{fig8}(b), both the $F_\text{max}$ and $F_\text{ave}$ tend to be steady-going when $\alpha>50\Gamma^2$ which makes the frequency sweeping range
cover all possible transitions with the sweeping period $T=4\Gamma^{-1}$ and the magnetic field strength $B=100~\text{G}$. The critical lower limit of the 
frequency ramp speed $\alpha$ can be resolved from Eq.(\ref{eq9}), while the upper limit depends on the adiabatic condition. For a fixed $\alpha=80\Gamma^2$ and 
$B=100~\text{G}$, with a longer sweeping period $T$, the coolable velocity region becomes wider according to Eq.(\ref{eq9}). For $T>2\Gamma^{-1}$ 
in Fig.\ref{fig8}(c), the average SWAP force $F_\text{ave}$ for velocities from $30\Gamma/k$ to $50\Gamma/k$ changes a little since the frequency 
range $\Delta_\text{T}$ is sufficient large. However, the case is different for small velocities as an approximately linear decreasement of the maximum cooling 
force $F_\text{max}$ by increasing the sweeping period $T$ is shown in Fig.\ref{fig8}(c). Since the frequency sweeping range already covers all the 
transitions at the small velocity region [$v<v_b$] for $T=2\Gamma^{-1}$, the momentum transfer $\delta p$ within one sweeping period become saturated even with 
lager $\Delta_\text{T}$. Consequently, the forces for small velocities $F=\delta p/T$ become smaller with a longer $T$. However, the period $T$ can not be 
drastically shortened, otherwise the population in excited state can not be deexcited back to the ground state at the beginning of each sweeping period, which 
makes the molecules be transferred away from zero momentum \cite{Muniz2018,Bartolotta2018}.
.

We have also checked the dependence of the SWAP force on the shift of the center frequency, as shown in Fig.\ref{fig8}(d). Within an interval from $-40\Gamma$ to 
$-10\Gamma$, the maximum cooling force slightly changes with considerable enhancement. The forces for large velocities show a similar steady-going behaviour. These
indicate that the long-term stablization of the cooling laser might not be strictly fulfilled for the SWAP schem, in contrast to the conventional Doppler cooling 
where the detuning should be precisely controlled at the magnitude of $\sim$MHz \cite{Wang2018}.

From above discussions, we conclude that the SWAP force is robust within a wide range of the SWAP parameters once the adiabatic condition and Eq.(\ref{eq9}) are
fulfilled. The maximum achievable cooling force $F_\text{max}$ is sensitive
to the energy gap between the exicited sublevels. The SWAP force always shows a weak velocity selective character, i.e., a rather wide
coolable velocity region up to $50\Gamma/k$ [$\sim 120~\text{m/s}$ for the BaF molecule] with typical SWAP parameters.

Let us consider the experimental realization of the large 
Rabi frequency $\Omega=50\Gamma$ and rapid sweeping with a ramp
speed of $\alpha=80\Gamma^2$. For narrow linewidth transitions, i.e., $\Gamma\sim \text{kHz}$, the
frequency sweeping can be easily achieved with an acousto-optic modulator \cite{Muniz2018,Petersen2018} and the 
laser intensity required is not too extreme. However, once extending to transitions with 
$\Gamma\sim\text{MHz}$, for example, the BaF molecule being investigated, the saturated intensity 
$I_s \sim 0.58~\text{mW/cm}^2$ and an $\Omega=50\Gamma$ indicates a high laser intensity of 
$\sim 3~\text{W/cm}^2$. On the other hand, a frequency ramp speed $\alpha=80\Gamma^2$ corresponds
to $\sim 4~\text{MHz/ns}$, which can be realized with an electro-optical modulator and the injection locking 
technique used in Refs.\cite{Teng2015,Kaufman2017} where a ramp speed of $\sim 100~\text{MHz/ns}$ was reported. 

\section{Resistance to leakage channels}\label{sec5}

\begin{figure}[b]
\includegraphics[width=0.48\textwidth]{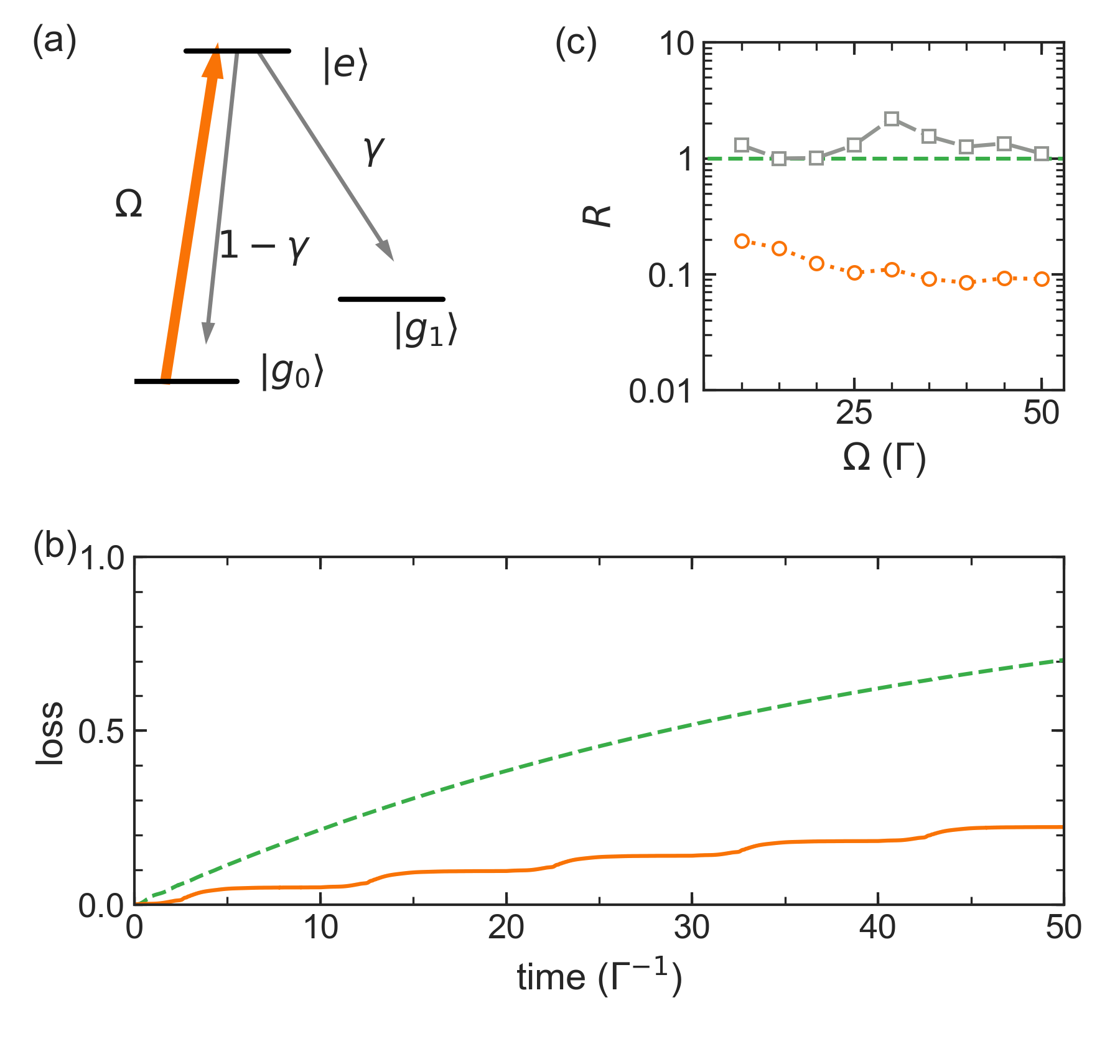}
\caption{\label{fig9}(Color online) The SWAP cooling effect on the leakage channels. (a) A schematic plot of 
a two-level system $|g_0\rangle\to|e\rangle$ driven by a pair of cooling laser beams (on-resonance Rabi 
frequency $\Omega$) with an additional leakage channel $|g_1\rangle$. The leakage rate is $\gamma$.
(b) The population loss to $|g_1\rangle$ within an evlution time of $50\Gamma^{-1}$. Initially, particles all 
populate the $|g_0\rangle$ state, and $\gamma=0.05$. For conventional Doppler cooling (green dashed line),
$\Omega=5\Gamma$, the detuning $\Delta=-kv$ with $v$ the velocity of the moving particle. The loss phenonmena
under the SWAP scheme ($\Omega=20\Gamma$, $\alpha=20\Gamma^2$, $T=10\Gamma^{-1}$) for 
a veloticy of $2\Gamma/k$ is different and shown in orange solid line.
(c) The dependence of the loss ratio $R$ of the SWAP cooling on the on-resonance Rabi frequency $
\Omega$ for the small ($v=2\Gamma/k$, orange circles) and large ($v=20\Gamma/k$, gray squares) velocities.
}
\end{figure}

For molecules, the closed cycling transitions generally can
not be perfectly realized due to the leakage channels, such as the higher vibrational states and the intermediate
$\Delta$ state \cite{Chen2016,Yeo2015}. Since the SWAP cooling scheme employs the stimulated emission to
transfer the excited particles back to the ground state, the fraction of spontaneous decay gets partially supressed. Hence,
compared to the conventional Doppler cooling, a smaller loss to the possible leakage channels might be expected.
In order to get some idea of such an effect, we consider a simple three-level system, i.e., a two-level transition driven by the frequency-chirped
cooling lasers with an additional loss channel; see Fig.\ref{fig9}(a). In the conventional Doppler cooling, a
loss rate $\gamma=0.05$ allows a maximum scattering photon number of $\gamma^{-1}\sim20$ before the particle
populates the leakage channel. As shown in Fig.\ref{fig9}(b), with an evolution time of $50\Gamma/k$, the 
loss fraction reaches $\sim 0.7$ after scattering $\sim 14$ photons. However, under the SWAP scheme,
with an equivalent evolution time, i.e., five sweeping periods, the loss $\ell_\text{swap}$ is
several times lower, and the increasing is nearly linear to the number
of the sweeping period before the cooling dies.

To quantitatively evaluate the resistance of the SWAP cooling to the leakage, we introduce a loss ratio defined by
\begin{equation}\label{eq10}
R = r_\text{swap}/r_\text{conv}.
\end{equation}
Here $r_\text{swap}$ indicates the population loss in the SWAP cooling once the moving particle changing its
momentum by $-\hbar k$, resulting a definition as $r_\text{swap}=\ell_\text{swap}\hbar k/\delta p_\text{swap}$
with $\delta p_\text{swap}(\tau)=|\int_0^\tau \langle F(t)\rangle dt|$ for an evolution time of $\tau$.
In the conventional Doppler cooling, $r_\text{conv}\sim \gamma$. Figure \ref{fig9}(c) shows the loss
ratios $R$ for two different velocities with various $\Omega$ under the SWAP scheme. For a large
velocity, for example, $v=20\Gamma/k$, the ratios are always larger than one, which means that the
cooling effect of the SWAP scheme is less significant than the conventional Doppler case as less momenta
($\sim 10$ photons for five periods) are transferred but the loss lies in a similar or larger level. 

However, for a small velocity, $v=2\Gamma/k$, the ratio is typically around $\sim 0.1$. 
This indicates that the leakage along with a decrease of the particle momentum by $\hbar k$ is one order
of magnitude smaller for the SWAP scheme than the Doppler cooling. The reason lies in that 
the SWAP cooling introduces more
than $2\hbar k$ momentum transfer during one sweeping period due to the nontrivial Bragg oscillations. This
is consistent with the effective condition $|kv|\ll\Omega$ where the Bragg oscillations work \cite{Bartolotta2018}. 
Meanwhile, with a better fulfillment of the adiabatic condition $\Omega^2\gg\alpha$
by increasing $\Omega$, the ratio $R$ decreases, as shown Fig.\ref{fig9}(c).
By recalling the discussion in Sec.\ref{sec3A}, the condition $v\ll \alpha\Gamma^{-1}/2k$ should be
fulfilled to depress the spontaneous decay and consequently achieve an effective SWAP cooling. We
claim here that, compared to the conventional Doppler cooling, a better resistance to the leakage channels in
the SWAP cooling occurs in the small velocity region, i.e., $v\ll\Omega/k$ and $v\ll \alpha\Gamma/2k$. 
This makes the SWAP cooling well-suited for particles with leakage channels. 

\section{Conclusion}

In summary, we have analyzed the feasibility of applying the SWAP cooling scheme for the multi-level type-II transitions
and further much more complex diatomic molecules. The angled magnetic field can not only remix the Zeeman dark 
states in the type-II transitions, but also introduce an enhancement to the SWAP cooling force at the small veloticy region. When 
the energy splittings between each two neighbouring sublevels from the magnetic field is larger than the Doppler shift, the multi-level
system can be decomposed into several two-level sub-systems in time ordering. Ideally, 
a momentum transfer of twice the photon momentum during one sweeping period is allowed and results in an enhanced cooling force appears for velocities below $v_b$.

Although the time order of the stimulated excitations and deexcitations in the SWAP scheme for the diatomic molecules
can not be perfectly guaranteed even with a large magnetic field of $B>100~\text{G}$, we still observe a 
$\sim 4\times$ enhancement of the maximum achievable cooling force in the small velocity region.
The cooling forces for a velocities larger than $\sim 100~\text{m/s}$ are still at the same magnitude with the maximum Doppler
cooling force. Different from the conventional cooling scheme, the SWAP scheme does not require a sideband modulation, and is less sensitive to the 
long-term stablization of the laser frequency. Such properties indicate a better experimental realization, and the
applications in laser slowing of a molecular beam might be possible as well. Finally, we have checked the resistance of the SWAP cooling
to the leakage channels, opening the door of laser cooling an ordinary molecule that lacks a closed-cycling
transition.

\begin{acknowledgments}

We acknowledge the support from the National Key Research and Development Program of China under Grant No.2018YFA0307200, National Natural Science Foundation of China under Grant No. 91636104, Natural Science Foundation of Zhejiang province under Grant No. LZ18A040001, and the Fundamental Research Funds for the Central Universities. 

\end{acknowledgments}

\appendix
\section{\label{appA} Derivation of the matrix elements of the electric dipole transitions under large magnetic field}

\begin{table}[]
\caption{\label{tabA}
Calculated matrix elements for electric dipole transitions from $\vert X, N=1,-\rangle$ state to $\vert A,J'=1/2,+\rangle$ state under a large magntic field.}
\begin{ruledtabular}
\begin{tabular}{ccc | rrrr}
~ & ~ & ~ & $m'_J=-1/2$ & $-1/2$ & $+1/2$ & $+1/2$ \\
$m_S$ & $m_N$ & $m_I$ & $m'_I=+1/2$ & $-1/2$ & $-1/2$ & $+1/2$ \\
\hline
$-1/2$ & $+1$ & $+1/2$ & 0.6084 & 0 & 0 & 0.1925 \\
~ & $+1$ & $-1/2$ & 0 & 0.6084 & 0.1925 & 0 \\
~ &  $0$ & $+1/2$ & -0.5443 & 0 & 0 & -0.1925 \\
~ &  $0$ & $-1/2$ & 0 & -0.5443 & -0.1925 & 0 \\
~ & $-1$ & $+1/2$ & 0.4082 & 0 & 0 & 0 \\
~ & $-1$ & $-1/2$ & 0 & 0.4082 & 0 & 0 \\
$+1/2$ & $-1$ & $-1/2$ & 0 & 0.1925 & 0.6084 & 0 \\
~ & $-1$ & $+1/2$ & 0.1925 & 0 & 0 & 0.6084 \\
~ &  $0$ & $-1/2$ & 0 & -0.1925 & -0.5443 & 0 \\
~ &  $0$ & $+1/2$ & -0.1925 & 0 & 0 & -0.5443 \\
~ & $+1$ & $-1/2$ & 0 & 0 & 0.4082 & 0 \\
~ & $+1$ & $+1/2$ & 0 & 0 & 0 & 0.4082 \\
\end{tabular}
\end{ruledtabular}
\end{table}

Following the labels used in Ref.\cite{Chen2016} and Eq.(6.149) in Ref.\cite{Brown2003}, the ground state in fully decoupled form can be written as 
\begin{eqnarray}
|g\rangle & = & |\Lambda; S,m_S; N,m_N; I,m_I\rangle \nonumber\\
& = & \sum\limits_{J,m_J}\sum\limits_{\Sigma} (-1)^{S-N+m_J+J+\Omega}\sqrt{(2J+1)(2N+1)} \nonumber\\
& & \times \left(\begin{array}{ccc} N & S & J \\ m_N & m_S & -m_J\end{array}\right)
\left(\begin{array}{ccc} N & S & J \\ \Lambda & \Sigma & -\Omega\end{array}\right) \nonumber\\
& &  \times|\Lambda;S,\Sigma;J,\Omega,m_J\rangle|I,m_I\rangle,
\end{eqnarray}
while the excited state is 
\begin{eqnarray}
|e\rangle & = & ||\Lambda'|; J', m'_J; I', m'_I; +\rangle \nonumber\\
& = & \frac{1}{\sqrt{2}}(|\Lambda';S',\Sigma';J',\Omega',m'_J\rangle + (-1)^{J'-S'} \nonumber\\
&&\times|-\Lambda';S',-\Sigma';J',-\Omega',m'_J\rangle)|I',m'_I\rangle.
\end{eqnarray}
Then, the matrix element for electric dipole transition from $|g\rangle$ to $|e\rangle$ is
\begin{eqnarray}\label{eqA3}
\langle d \rangle &=& \langle e|T_p^1(\hat{d})|g\rangle \nonumber \\
&=& \frac{1}{\sqrt{2}}\delta_{I,I'}\delta_{m_I,m'_I}\sum\limits_{J,m_J}\sum\limits_{\Sigma} (-1)^{S-N+m_J+J+\Omega} \nonumber \\
&\times& \sqrt{(2J+1)(2N+1)} \left(\begin{array}{ccc} N & S & J \\ m_N & m_S & -m_J\end{array}\right)
\left(\begin{array}{ccc} N & S & J \\ \Lambda & \Sigma & -\Omega\end{array}\right) \nonumber\\
&\times& ( \langle \Lambda';S',\Sigma';J',\Omega',m'_J|T_p^1(\hat{d})|\Lambda;S,\Sigma;J,\Omega,m_J\rangle +\nonumber\\
&& \langle -\Lambda';S',-\Sigma';J',-\Omega',m'_J|T_p^1(\hat{d})|\Lambda;S,\Sigma;J,\Omega,m_J\rangle ).\nonumber\\
\end{eqnarray}
By applying Wigner-Eckart theorem, the internal term
\begin{eqnarray}
&&\langle \Lambda';S',\Sigma';J',\Omega',m'_J|T_p^1(\hat{d})|\Lambda;S,\Sigma;J,\Omega,m_J\rangle \nonumber\\
&&~~~~=\delta_{\Sigma,\Sigma'}\sum\limits_{q}(-1)^{m'_J-\Omega'}\sqrt{(2J'+1)(2J+1)} \nonumber\\
&&~~~~\times \left(\begin{array}{ccc} J' & 1 & J \\ -m'_J & p & m_J\end{array}\right)
\left(\begin{array}{ccc} J' & 1 & J \\ -\Omega' & q & \Omega\end{array}\right) 
\langle \Lambda'|T_q^1(\hat{d})|\Lambda\rangle. \nonumber\\
\end{eqnarray}
Since matrix element $\langle\Lambda'|T_q^1(\hat{d})|\Lambda\rangle$ is common for all $\Delta\Lambda=\pm 1$ transitions, we obtain from
Eq.(\ref{eqA3}) all possible hyperfine transition strength, i.e., values of $A$ in Eq.(\ref{eq3}), in $|X,N=1,-\rangle \to |A,J=1/2,+\rangle$ transtions 
under large magnetic field, as listed in Table \ref{tabA}.

\bibliographystyle{apsrev4-1}

\bibliography{swap_cooling_molecule}

\begin{thebibliography}{40}%
\makeatletter
\providecommand \@ifxundefined [1]{%
 \@ifx{#1\undefined}
}%
\providecommand \@ifnum [1]{%
 \ifnum #1\expandafter \@firstoftwo
 \else \expandafter \@secondoftwo
 \fi
}%
\providecommand \@ifx [1]{%
 \ifx #1\expandafter \@firstoftwo
 \else \expandafter \@secondoftwo
 \fi
}%
\providecommand \natexlab [1]{#1}%
\providecommand \enquote  [1]{``#1''}%
\providecommand \bibnamefont  [1]{#1}%
\providecommand \bibfnamefont [1]{#1}%
\providecommand \citenamefont [1]{#1}%
\providecommand \href@noop [0]{\@secondoftwo}%
\providecommand \href [0]{\begingroup \@sanitize@url \@href}%
\providecommand \@href[1]{\@@startlink{#1}\@@href}%
\providecommand \@@href[1]{\endgroup#1\@@endlink}%
\providecommand \@sanitize@url [0]{\catcode `\\12\catcode `\$12\catcode
  `\&12\catcode `\#12\catcode `\^12\catcode `\_12\catcode `\%12\relax}%
\providecommand \@@startlink[1]{}%
\providecommand \@@endlink[0]{}%
\providecommand \url  [0]{\begingroup\@sanitize@url \@url }%
\providecommand \@url [1]{\endgroup\@href {#1}{\urlprefix }}%
\providecommand \urlprefix  [0]{URL }%
\providecommand \Eprint [0]{\href }%
\providecommand \doibase [0]{http://dx.doi.org/}%
\providecommand \selectlanguage [0]{\@gobble}%
\providecommand \bibinfo  [0]{\@secondoftwo}%
\providecommand \bibfield  [0]{\@secondoftwo}%
\providecommand \translation [1]{[#1]}%
\providecommand \BibitemOpen [0]{}%
\providecommand \bibitemStop [0]{}%
\providecommand \bibitemNoStop [0]{.\EOS\space}%
\providecommand \EOS [0]{\spacefactor3000\relax}%
\providecommand \BibitemShut  [1]{\csname bibitem#1\endcsname}%
\let\auto@bib@innerbib\@empty
\bibitem [{\citenamefont {Chu}(1998)}]{Chu1998}%
  \BibitemOpen
  \bibfield  {author} {\bibinfo {author} {\bibfnamefont {S.}~\bibnamefont
  {Chu}},\ }\href {\doibase 10.1103/RevModPhys.70.685} {\bibfield  {journal}
  {\bibinfo  {journal} {Rev. Mod. Phys.}\ }\textbf {\bibinfo {volume} {70}},\
  \bibinfo {pages} {685} (\bibinfo {year} {1998})}\BibitemShut {NoStop}%
\bibitem [{\citenamefont {Phillips}(1998)}]{Phillips1998}%
  \BibitemOpen
  \bibfield  {author} {\bibinfo {author} {\bibfnamefont {W.~D.}\ \bibnamefont
  {Phillips}},\ }\href {\doibase 10.1103/RevModPhys.70.721} {\bibfield
  {journal} {\bibinfo  {journal} {Rev. Mod. Phys.}\ }\textbf {\bibinfo {volume}
  {70}},\ \bibinfo {pages} {721} (\bibinfo {year} {1998})}\BibitemShut
  {NoStop}%
\bibitem [{\citenamefont {Cohen-Tannoudji}(1998)}]{Cohen-Tannoudji1998}%
  \BibitemOpen
  \bibfield  {author} {\bibinfo {author} {\bibfnamefont {C.~N.}\ \bibnamefont
  {Cohen-Tannoudji}},\ }\href {\doibase 10.1103/RevModPhys.70.707} {\bibfield
  {journal} {\bibinfo  {journal} {Rev. Mod. Phys.}\ }\textbf {\bibinfo {volume}
  {70}},\ \bibinfo {pages} {707} (\bibinfo {year} {1998})}\BibitemShut
  {NoStop}%
\bibitem [{\citenamefont {Cornell}\ and\ \citenamefont
  {Wieman}(2002)}]{Cornell2002}%
  \BibitemOpen
  \bibfield  {author} {\bibinfo {author} {\bibfnamefont {E.~A.}\ \bibnamefont
  {Cornell}}\ and\ \bibinfo {author} {\bibfnamefont {C.~E.}\ \bibnamefont
  {Wieman}},\ }\href {\doibase 10.1103/RevModPhys.74.875} {\bibfield  {journal}
  {\bibinfo  {journal} {Rev. Mod. Phys.}\ }\textbf {\bibinfo {volume} {74}},\
  \bibinfo {pages} {875} (\bibinfo {year} {2002})}\BibitemShut {NoStop}%
\bibitem [{\citenamefont {Bloch}\ \emph {et~al.}(2008)\citenamefont {Bloch},
  \citenamefont {Dalibard},\ and\ \citenamefont {Zwerger}}]{Bloch2008}%
  \BibitemOpen
  \bibfield  {author} {\bibinfo {author} {\bibfnamefont {I.}~\bibnamefont
  {Bloch}}, \bibinfo {author} {\bibfnamefont {J.}~\bibnamefont {Dalibard}}, \
  and\ \bibinfo {author} {\bibfnamefont {W.}~\bibnamefont {Zwerger}},\ }\href
  {\doibase 10.1103/RevModPhys.80.885} {\bibfield  {journal} {\bibinfo
  {journal} {Rev. Mod. Phys.}\ }\textbf {\bibinfo {volume} {80}},\ \bibinfo
  {pages} {885} (\bibinfo {year} {2008})}\BibitemShut {NoStop}%
\bibitem [{\citenamefont {Ludlow}\ \emph {et~al.}(2015)\citenamefont {Ludlow},
  \citenamefont {Boyd}, \citenamefont {Ye}, \citenamefont {Peik},\ and\
  \citenamefont {Schmidt}}]{Ludlow2015}%
  \BibitemOpen
  \bibfield  {author} {\bibinfo {author} {\bibfnamefont {A.~D.}\ \bibnamefont
  {Ludlow}}, \bibinfo {author} {\bibfnamefont {M.~M.}\ \bibnamefont {Boyd}},
  \bibinfo {author} {\bibfnamefont {J.}~\bibnamefont {Ye}}, \bibinfo {author}
  {\bibfnamefont {E.}~\bibnamefont {Peik}}, \ and\ \bibinfo {author}
  {\bibfnamefont {P.~O.}\ \bibnamefont {Schmidt}},\ }\href {\doibase
  10.1103/RevModPhys.87.637} {\bibfield  {journal} {\bibinfo  {journal} {Rev.
  Mod. Phys.}\ }\textbf {\bibinfo {volume} {87}},\ \bibinfo {pages} {637}
  (\bibinfo {year} {2015})}\BibitemShut {NoStop}%
\bibitem [{\citenamefont {Bohn}\ \emph {et~al.}(2017)\citenamefont {Bohn},
  \citenamefont {Rey},\ and\ \citenamefont {Ye}}]{Bohn2017}%
  \BibitemOpen
  \bibfield  {author} {\bibinfo {author} {\bibfnamefont {J.~L.}\ \bibnamefont
  {Bohn}}, \bibinfo {author} {\bibfnamefont {A.~M.}\ \bibnamefont {Rey}}, \
  and\ \bibinfo {author} {\bibfnamefont {J.}~\bibnamefont {Ye}},\ }\href
  {\doibase 10.1126/science.aam6299} {\bibfield  {journal} {\bibinfo  {journal}
  {Science}\ }\textbf {\bibinfo {volume} {357}},\ \bibinfo {pages} {1002}
  (\bibinfo {year} {2017})}\BibitemShut {NoStop}%
\bibitem [{\citenamefont {Safronova}\ \emph {et~al.}(2018)\citenamefont
  {Safronova}, \citenamefont {Budker}, \citenamefont {DeMille}, \citenamefont
  {Kimball}, \citenamefont {Derevianko},\ and\ \citenamefont
  {Clark}}]{Safronova2018}%
  \BibitemOpen
  \bibfield  {author} {\bibinfo {author} {\bibfnamefont {M.~S.}\ \bibnamefont
  {Safronova}}, \bibinfo {author} {\bibfnamefont {D.}~\bibnamefont {Budker}},
  \bibinfo {author} {\bibfnamefont {D.}~\bibnamefont {DeMille}}, \bibinfo
  {author} {\bibfnamefont {D.~F.~J.}\ \bibnamefont {Kimball}}, \bibinfo
  {author} {\bibfnamefont {A.}~\bibnamefont {Derevianko}}, \ and\ \bibinfo
  {author} {\bibfnamefont {C.~W.}\ \bibnamefont {Clark}},\ }\href {\doibase
  10.1103/RevModPhys.90.025008} {\bibfield  {journal} {\bibinfo  {journal}
  {Rev. Mod. Phys.}\ }\textbf {\bibinfo {volume} {90}},\ \bibinfo {pages}
  {025008} (\bibinfo {year} {2018})}\BibitemShut {NoStop}%
\bibitem [{\citenamefont {Metcalf}(2017)}]{Metcalf2017}%
  \BibitemOpen
  \bibfield  {author} {\bibinfo {author} {\bibfnamefont {H.}~\bibnamefont
  {Metcalf}},\ }\href {\doibase 10.1103/RevModPhys.89.041001} {\bibfield
  {journal} {\bibinfo  {journal} {Rev. Mod. Phys.}\ }\textbf {\bibinfo {volume}
  {89}},\ \bibinfo {pages} {041001} (\bibinfo {year} {2017})}\BibitemShut
  {NoStop}%
\bibitem [{\citenamefont {Lu}\ \emph {et~al.}(2005)\citenamefont {Lu},
  \citenamefont {Miao},\ and\ \citenamefont {Metcalf}}]{Lu2005}%
  \BibitemOpen
  \bibfield  {author} {\bibinfo {author} {\bibfnamefont {T.}~\bibnamefont
  {Lu}}, \bibinfo {author} {\bibfnamefont {X.}~\bibnamefont {Miao}}, \ and\
  \bibinfo {author} {\bibfnamefont {H.}~\bibnamefont {Metcalf}},\ }\href
  {\doibase 10.1103/PhysRevA.71.061405} {\bibfield  {journal} {\bibinfo
  {journal} {Phys. Rev. A}\ }\textbf {\bibinfo {volume} {71}},\ \bibinfo
  {pages} {061405} (\bibinfo {year} {2005})}\BibitemShut {NoStop}%
\bibitem [{\citenamefont {Miao}\ \emph {et~al.}(2007)\citenamefont {Miao},
  \citenamefont {Wertz}, \citenamefont {Cohen},\ and\ \citenamefont
  {Metcalf}}]{Miao2007}%
  \BibitemOpen
  \bibfield  {author} {\bibinfo {author} {\bibfnamefont {X.}~\bibnamefont
  {Miao}}, \bibinfo {author} {\bibfnamefont {E.}~\bibnamefont {Wertz}},
  \bibinfo {author} {\bibfnamefont {M.~G.}\ \bibnamefont {Cohen}}, \ and\
  \bibinfo {author} {\bibfnamefont {H.}~\bibnamefont {Metcalf}},\ }\href
  {\doibase 10.1103/PhysRevA.75.011402} {\bibfield  {journal} {\bibinfo
  {journal} {Phys. Rev. A}\ }\textbf {\bibinfo {volume} {75}},\ \bibinfo
  {pages} {011402} (\bibinfo {year} {2007})}\BibitemShut {NoStop}%
\bibitem [{\citenamefont {Jayich}\ \emph {et~al.}(2014)\citenamefont {Jayich},
  \citenamefont {Vutha}, \citenamefont {Hummon}, \citenamefont {Porto},\ and\
  \citenamefont {Campbell}}]{Jayich2014}%
  \BibitemOpen
  \bibfield  {author} {\bibinfo {author} {\bibfnamefont {A.~M.}\ \bibnamefont
  {Jayich}}, \bibinfo {author} {\bibfnamefont {A.~C.}\ \bibnamefont {Vutha}},
  \bibinfo {author} {\bibfnamefont {M.~T.}\ \bibnamefont {Hummon}}, \bibinfo
  {author} {\bibfnamefont {J.~V.}\ \bibnamefont {Porto}}, \ and\ \bibinfo
  {author} {\bibfnamefont {W.~C.}\ \bibnamefont {Campbell}},\ }\href {\doibase
  10.1103/PhysRevA.89.023425} {\bibfield  {journal} {\bibinfo  {journal} {Phys.
  Rev. A}\ }\textbf {\bibinfo {volume} {89}},\ \bibinfo {pages} {023425}
  (\bibinfo {year} {2014})}\BibitemShut {NoStop}%
\bibitem [{\citenamefont {S\"oding}\ \emph {et~al.}(1997)\citenamefont
  {S\"oding}, \citenamefont {Grimm}, \citenamefont {Ovchinnikov}, \citenamefont
  {Bouyer},\ and\ \citenamefont {Salomon}}]{Soeding1997}%
  \BibitemOpen
  \bibfield  {author} {\bibinfo {author} {\bibfnamefont {J.}~\bibnamefont
  {S\"oding}}, \bibinfo {author} {\bibfnamefont {R.}~\bibnamefont {Grimm}},
  \bibinfo {author} {\bibfnamefont {Y.~B.}\ \bibnamefont {Ovchinnikov}},
  \bibinfo {author} {\bibfnamefont {P.}~\bibnamefont {Bouyer}}, \ and\ \bibinfo
  {author} {\bibfnamefont {C.}~\bibnamefont {Salomon}},\ }\href {\doibase
  10.1103/PhysRevLett.78.1420} {\bibfield  {journal} {\bibinfo  {journal}
  {Phys. Rev. Lett.}\ }\textbf {\bibinfo {volume} {78}},\ \bibinfo {pages}
  {1420} (\bibinfo {year} {1997})}\BibitemShut {NoStop}%
\bibitem [{\citenamefont {Yatsenko}\ and\ \citenamefont
  {Metcalf}(2004)}]{Yatsenko2004}%
  \BibitemOpen
  \bibfield  {author} {\bibinfo {author} {\bibfnamefont {L.}~\bibnamefont
  {Yatsenko}}\ and\ \bibinfo {author} {\bibfnamefont {H.}~\bibnamefont
  {Metcalf}},\ }\href {\doibase 10.1103/PhysRevA.70.063402} {\bibfield
  {journal} {\bibinfo  {journal} {Phys. Rev. A}\ }\textbf {\bibinfo {volume}
  {70}},\ \bibinfo {pages} {063402} (\bibinfo {year} {2004})}\BibitemShut
  {NoStop}%
\bibitem [{\citenamefont {Partlow}\ \emph {et~al.}(2004)\citenamefont
  {Partlow}, \citenamefont {Miao}, \citenamefont {Bochmann}, \citenamefont
  {Cashen},\ and\ \citenamefont {Metcalf}}]{Partlow2004}%
  \BibitemOpen
  \bibfield  {author} {\bibinfo {author} {\bibfnamefont {M.}~\bibnamefont
  {Partlow}}, \bibinfo {author} {\bibfnamefont {X.}~\bibnamefont {Miao}},
  \bibinfo {author} {\bibfnamefont {J.}~\bibnamefont {Bochmann}}, \bibinfo
  {author} {\bibfnamefont {M.}~\bibnamefont {Cashen}}, \ and\ \bibinfo {author}
  {\bibfnamefont {H.}~\bibnamefont {Metcalf}},\ }\href {\doibase
  10.1103/PhysRevLett.93.213004} {\bibfield  {journal} {\bibinfo  {journal}
  {Phys. Rev. Lett.}\ }\textbf {\bibinfo {volume} {93}},\ \bibinfo {pages}
  {213004} (\bibinfo {year} {2004})}\BibitemShut {NoStop}%
\bibitem [{\citenamefont {Corder}\ \emph {et~al.}(2015)\citenamefont {Corder},
  \citenamefont {Arnold},\ and\ \citenamefont {Metcalf}}]{Corder2015}%
  \BibitemOpen
  \bibfield  {author} {\bibinfo {author} {\bibfnamefont {C.}~\bibnamefont
  {Corder}}, \bibinfo {author} {\bibfnamefont {B.}~\bibnamefont {Arnold}}, \
  and\ \bibinfo {author} {\bibfnamefont {H.}~\bibnamefont {Metcalf}},\ }\href
  {\doibase 10.1103/PhysRevLett.114.043002} {\bibfield  {journal} {\bibinfo
  {journal} {Phys. Rev. Lett.}\ }\textbf {\bibinfo {volume} {114}},\ \bibinfo
  {pages} {043002} (\bibinfo {year} {2015})}\BibitemShut {NoStop}%
\bibitem [{\citenamefont {Shuman}\ \emph {et~al.}(2010)\citenamefont {Shuman},
  \citenamefont {Barry},\ and\ \citenamefont {DeMille}}]{Shuman2010}%
  \BibitemOpen
  \bibfield  {author} {\bibinfo {author} {\bibfnamefont {E.~S.}\ \bibnamefont
  {Shuman}}, \bibinfo {author} {\bibfnamefont {J.~F.}\ \bibnamefont {Barry}}, \
  and\ \bibinfo {author} {\bibfnamefont {D.}~\bibnamefont {DeMille}},\ }\href
  {\doibase https://doi.org/10.1038/nature09443} {\bibfield  {journal}
  {\bibinfo  {journal} {Nature}\ }\textbf {\bibinfo {volume} {467}},\ \bibinfo
  {pages} {820} (\bibinfo {year} {2010})}\BibitemShut {NoStop}%
\bibitem [{\citenamefont {Hummon}\ \emph {et~al.}(2013)\citenamefont {Hummon},
  \citenamefont {Yeo}, \citenamefont {Stuhl}, \citenamefont {Collopy},
  \citenamefont {Xia},\ and\ \citenamefont {Ye}}]{Hummon2013}%
  \BibitemOpen
  \bibfield  {author} {\bibinfo {author} {\bibfnamefont {M.~T.}\ \bibnamefont
  {Hummon}}, \bibinfo {author} {\bibfnamefont {M.}~\bibnamefont {Yeo}},
  \bibinfo {author} {\bibfnamefont {B.~K.}\ \bibnamefont {Stuhl}}, \bibinfo
  {author} {\bibfnamefont {A.~L.}\ \bibnamefont {Collopy}}, \bibinfo {author}
  {\bibfnamefont {Y.}~\bibnamefont {Xia}}, \ and\ \bibinfo {author}
  {\bibfnamefont {J.}~\bibnamefont {Ye}},\ }\href {\doibase
  10.1103/PhysRevLett.110.143001} {\bibfield  {journal} {\bibinfo  {journal}
  {Phys. Rev. Lett.}\ }\textbf {\bibinfo {volume} {110}},\ \bibinfo {pages}
  {143001} (\bibinfo {year} {2013})}\BibitemShut {NoStop}%
\bibitem [{\citenamefont {Di~Rosa}(2004)}]{DiRosa2004}%
  \BibitemOpen
  \bibfield  {author} {\bibinfo {author} {\bibfnamefont {M.~D.}\ \bibnamefont
  {Di~Rosa}},\ }\href {\doibase 10.1140/epjd/e2004-00167-2} {\bibfield
  {journal} {\bibinfo  {journal} {The European Physical Journal D - Atomic,
  Molecular, Optical and Plasma Physics}\ }\textbf {\bibinfo {volume} {31}},\
  \bibinfo {pages} {395} (\bibinfo {year} {2004})}\BibitemShut {NoStop}%
\bibitem [{\citenamefont {Chen}\ \emph {et~al.}(2016)\citenamefont {Chen},
  \citenamefont {Bu},\ and\ \citenamefont {Yan}}]{Chen2016}%
  \BibitemOpen
  \bibfield  {author} {\bibinfo {author} {\bibfnamefont {T.}~\bibnamefont
  {Chen}}, \bibinfo {author} {\bibfnamefont {W.}~\bibnamefont {Bu}}, \ and\
  \bibinfo {author} {\bibfnamefont {B.}~\bibnamefont {Yan}},\ }\href {\doibase
  10.1103/PhysRevA.94.063415} {\bibfield  {journal} {\bibinfo  {journal} {Phys.
  Rev. A}\ }\textbf {\bibinfo {volume} {94}},\ \bibinfo {pages} {063415}
  (\bibinfo {year} {2016})}\BibitemShut {NoStop}%
\bibitem [{\citenamefont {Stuhl}\ \emph {et~al.}(2008)\citenamefont {Stuhl},
  \citenamefont {Sawyer}, \citenamefont {Wang},\ and\ \citenamefont
  {Ye}}]{Stuhl2008}%
  \BibitemOpen
  \bibfield  {author} {\bibinfo {author} {\bibfnamefont {B.~K.}\ \bibnamefont
  {Stuhl}}, \bibinfo {author} {\bibfnamefont {B.~C.}\ \bibnamefont {Sawyer}},
  \bibinfo {author} {\bibfnamefont {D.}~\bibnamefont {Wang}}, \ and\ \bibinfo
  {author} {\bibfnamefont {J.}~\bibnamefont {Ye}},\ }\href {\doibase
  10.1103/PhysRevLett.101.243002} {\bibfield  {journal} {\bibinfo  {journal}
  {Phys. Rev. Lett.}\ }\textbf {\bibinfo {volume} {101}},\ \bibinfo {pages}
  {243002} (\bibinfo {year} {2008})}\BibitemShut {NoStop}%
\bibitem [{\citenamefont {Chen}\ \emph {et~al.}(2017)\citenamefont {Chen},
  \citenamefont {Bu},\ and\ \citenamefont {Yan}}]{Chen2017}%
  \BibitemOpen
  \bibfield  {author} {\bibinfo {author} {\bibfnamefont {T.}~\bibnamefont
  {Chen}}, \bibinfo {author} {\bibfnamefont {W.}~\bibnamefont {Bu}}, \ and\
  \bibinfo {author} {\bibfnamefont {B.}~\bibnamefont {Yan}},\ }\href {\doibase
  10.1103/PhysRevA.96.053401} {\bibfield  {journal} {\bibinfo  {journal} {Phys.
  Rev. A}\ }\textbf {\bibinfo {volume} {96}},\ \bibinfo {pages} {053401}
  (\bibinfo {year} {2017})}\BibitemShut {NoStop}%
\bibitem [{\citenamefont {Kozyryev}\ \emph {et~al.}(2018)\citenamefont
  {Kozyryev}, \citenamefont {Baum}, \citenamefont {Aldridge}, \citenamefont
  {Yu}, \citenamefont {Eyler},\ and\ \citenamefont {Doyle}}]{Kozyryev2018}%
  \BibitemOpen
  \bibfield  {author} {\bibinfo {author} {\bibfnamefont {I.}~\bibnamefont
  {Kozyryev}}, \bibinfo {author} {\bibfnamefont {L.}~\bibnamefont {Baum}},
  \bibinfo {author} {\bibfnamefont {L.}~\bibnamefont {Aldridge}}, \bibinfo
  {author} {\bibfnamefont {P.}~\bibnamefont {Yu}}, \bibinfo {author}
  {\bibfnamefont {E.~E.}\ \bibnamefont {Eyler}}, \ and\ \bibinfo {author}
  {\bibfnamefont {J.~M.}\ \bibnamefont {Doyle}},\ }\href {\doibase
  10.1103/PhysRevLett.120.063205} {\bibfield  {journal} {\bibinfo  {journal}
  {Phys. Rev. Lett.}\ }\textbf {\bibinfo {volume} {120}},\ \bibinfo {pages}
  {063205} (\bibinfo {year} {2018})}\BibitemShut {NoStop}%
\bibitem [{\citenamefont {Metcalf}\ and\ \citenamefont {der
  Straten}(1999)}]{Metcalf1999}%
  \BibitemOpen
  \bibfield  {author} {\bibinfo {author} {\bibfnamefont {H.}~\bibnamefont
  {Metcalf}}\ and\ \bibinfo {author} {\bibfnamefont {P.~V.}\ \bibnamefont {der
  Straten}},\ }\href@noop {} {\emph {\bibinfo {title} {Laser cooling and
  trapping}}}\ (\bibinfo  {publisher} {Springer},\ \bibinfo {year}
  {1999})\BibitemShut {NoStop}%
\bibitem [{\citenamefont {Dalibard}\ and\ \citenamefont
  {Tannoudji}(1989)}]{DALIBARD1989}%
  \BibitemOpen
  \bibfield  {author} {\bibinfo {author} {\bibfnamefont {J.}~\bibnamefont
  {Dalibard}}\ and\ \bibinfo {author} {\bibfnamefont {C.}~\bibnamefont
  {Tannoudji}},\ }\href {\doibase 10.1364/JOSAB.6.002023} {\bibfield  {journal}
  {\bibinfo  {journal} {J. Opt. Soc. Am. B}\ }\textbf {\bibinfo {volume} {6}},\
  \bibinfo {pages} {2023} (\bibinfo {year} {1989})}\BibitemShut {NoStop}%
\bibitem [{\citenamefont {Ungar}\ \emph {et~al.}(1989)\citenamefont {Ungar},
  \citenamefont {Weiss}, \citenamefont {Riis},\ and\ \citenamefont
  {Chu}}]{UNGAR1989}%
  \BibitemOpen
  \bibfield  {author} {\bibinfo {author} {\bibfnamefont {P.}~\bibnamefont
  {Ungar}}, \bibinfo {author} {\bibfnamefont {D.}~\bibnamefont {Weiss}},
  \bibinfo {author} {\bibfnamefont {E.}~\bibnamefont {Riis}}, \ and\ \bibinfo
  {author} {\bibfnamefont {S.}~\bibnamefont {Chu}},\ }\href {\doibase
  10.1364/JOSAB.6.002058} {\bibfield  {journal} {\bibinfo  {journal} {J. Opt.
  Soc. Am. B}\ }\textbf {\bibinfo {volume} {6}},\ \bibinfo {pages} {2058}
  (\bibinfo {year} {1989})}\BibitemShut {NoStop}%
\bibitem [{\citenamefont {Norcia}\ \emph {et~al.}(2018)\citenamefont {Norcia},
  \citenamefont {Cline}, \citenamefont {Bartolotta}, \citenamefont {Holland},\
  and\ \citenamefont {Thompson}}]{Norcia2018}%
  \BibitemOpen
  \bibfield  {author} {\bibinfo {author} {\bibfnamefont {M.~A.}\ \bibnamefont
  {Norcia}}, \bibinfo {author} {\bibfnamefont {J.~R.~K.}\ \bibnamefont
  {Cline}}, \bibinfo {author} {\bibfnamefont {J.~P.}\ \bibnamefont
  {Bartolotta}}, \bibinfo {author} {\bibfnamefont {M.~J.}\ \bibnamefont
  {Holland}}, \ and\ \bibinfo {author} {\bibfnamefont {J.~K.}\ \bibnamefont
  {Thompson}},\ }\href {http://stacks.iop.org/1367-2630/20/i=2/a=023021}
  {\bibfield  {journal} {\bibinfo  {journal} {New Journal of Physics}\ }\textbf
  {\bibinfo {volume} {20}},\ \bibinfo {pages} {023021} (\bibinfo {year}
  {2018})}\BibitemShut {NoStop}%
\bibitem [{\citenamefont {Muniz}\ \emph {et~al.}(2018)\citenamefont {Muniz},
  \citenamefont {Norcia}, \citenamefont {Cline},\ and\ \citenamefont
  {Thompson}}]{Muniz2018}%
  \BibitemOpen
  \bibfield  {author} {\bibinfo {author} {\bibfnamefont {J.~A.}\ \bibnamefont
  {Muniz}}, \bibinfo {author} {\bibfnamefont {M.~A.}\ \bibnamefont {Norcia}},
  \bibinfo {author} {\bibfnamefont {J.~R.~K.}\ \bibnamefont {Cline}}, \ and\
  \bibinfo {author} {\bibfnamefont {J.~K.}\ \bibnamefont {Thompson}},\ }\href
  {https://arxiv.org/abs/1806.00838} {\bibfield  {journal} {\bibinfo  {journal}
  {arXiv: 1806.00838}\ } (\bibinfo {year} {2018})}\BibitemShut {NoStop}%
\bibitem [{\citenamefont {Petersen}\ \emph {et~al.}(2018)\citenamefont
  {Petersen}, \citenamefont {Mühlbauer}, \citenamefont {Bougas}, \citenamefont
  {Sharma}, \citenamefont {Budker},\ and\ \citenamefont
  {Windpassinger}}]{Petersen2018}%
  \BibitemOpen
  \bibfield  {author} {\bibinfo {author} {\bibfnamefont {N.}~\bibnamefont
  {Petersen}}, \bibinfo {author} {\bibfnamefont {F.}~\bibnamefont
  {Mühlbauer}}, \bibinfo {author} {\bibfnamefont {L.}~\bibnamefont {Bougas}},
  \bibinfo {author} {\bibfnamefont {A.}~\bibnamefont {Sharma}}, \bibinfo
  {author} {\bibfnamefont {D.}~\bibnamefont {Budker}}, \ and\ \bibinfo {author}
  {\bibfnamefont {P.}~\bibnamefont {Windpassinger}},\ }\href
  {https://arxiv.org/abs/1809.06423v1} {\bibfield  {journal} {\bibinfo
  {journal} {arXiv: 1809.06423}\ } (\bibinfo {year} {2018})}\BibitemShut
  {NoStop}%
\bibitem [{\citenamefont {Greve}\ \emph {et~al.}(2018)\citenamefont {Greve},
  \citenamefont {Wu},\ and\ \citenamefont {Thompson}}]{Greve2018}%
  \BibitemOpen
  \bibfield  {author} {\bibinfo {author} {\bibfnamefont {G.~P.}\ \bibnamefont
  {Greve}}, \bibinfo {author} {\bibfnamefont {B.~C.}\ \bibnamefont {Wu}}, \
  and\ \bibinfo {author} {\bibfnamefont {J.~K.}\ \bibnamefont {Thompson}},\
  }\href {https://arxiv.org/abs/1805.04452} {\bibfield  {journal} {\bibinfo
  {journal} {arXiv:1805.04452}\ } (\bibinfo {year} {2018})}\BibitemShut
  {NoStop}%
\bibitem [{\citenamefont {Bartolotta}\ \emph {et~al.}(2018)\citenamefont
  {Bartolotta}, \citenamefont {Norcia}, \citenamefont {Cline}, \citenamefont
  {Thompson},\ and\ \citenamefont {Holland}}]{Bartolotta2018}%
  \BibitemOpen
  \bibfield  {author} {\bibinfo {author} {\bibfnamefont {J.~P.}\ \bibnamefont
  {Bartolotta}}, \bibinfo {author} {\bibfnamefont {M.~A.}\ \bibnamefont
  {Norcia}}, \bibinfo {author} {\bibfnamefont {J.~R.~K.}\ \bibnamefont
  {Cline}}, \bibinfo {author} {\bibfnamefont {J.~K.}\ \bibnamefont {Thompson}},
  \ and\ \bibinfo {author} {\bibfnamefont {M.~J.}\ \bibnamefont {Holland}},\
  }\href {\doibase 10.1103/PhysRevA.98.023404} {\bibfield  {journal} {\bibinfo
  {journal} {Phys. Rev. A}\ }\textbf {\bibinfo {volume} {98}},\ \bibinfo
  {pages} {023404} (\bibinfo {year} {2018})}\BibitemShut {NoStop}%
\bibitem [{\citenamefont {Oien}\ \emph {et~al.}(1997)\citenamefont {Oien},
  \citenamefont {McKinnie}, \citenamefont {Manson}, \citenamefont {Sandle},\
  and\ \citenamefont {Warrington}}]{Oien1997}%
  \BibitemOpen
  \bibfield  {author} {\bibinfo {author} {\bibfnamefont {A.~M.~L.}\
  \bibnamefont {Oien}}, \bibinfo {author} {\bibfnamefont {I.~T.}\ \bibnamefont
  {McKinnie}}, \bibinfo {author} {\bibfnamefont {P.~J.}\ \bibnamefont
  {Manson}}, \bibinfo {author} {\bibfnamefont {W.~J.}\ \bibnamefont {Sandle}},
  \ and\ \bibinfo {author} {\bibfnamefont {D.~M.}\ \bibnamefont {Warrington}},\
  }\href {\doibase 10.1103/PhysRevA.55.4621} {\bibfield  {journal} {\bibinfo
  {journal} {Phys. Rev. A}\ }\textbf {\bibinfo {volume} {55}},\ \bibinfo
  {pages} {4621} (\bibinfo {year} {1997})}\BibitemShut {NoStop}%
\bibitem [{\citenamefont {Tiwari}\ \emph {et~al.}(2008)\citenamefont {Tiwari},
  \citenamefont {Singh}, \citenamefont {Rawat},\ and\ \citenamefont
  {Mehendale}}]{Tiwari2008}%
  \BibitemOpen
  \bibfield  {author} {\bibinfo {author} {\bibfnamefont {V.~B.}\ \bibnamefont
  {Tiwari}}, \bibinfo {author} {\bibfnamefont {S.}~\bibnamefont {Singh}},
  \bibinfo {author} {\bibfnamefont {H.~S.}\ \bibnamefont {Rawat}}, \ and\
  \bibinfo {author} {\bibfnamefont {S.~C.}\ \bibnamefont {Mehendale}},\ }\href
  {\doibase 10.1103/PhysRevA.78.063421} {\bibfield  {journal} {\bibinfo
  {journal} {Phys. Rev. A}\ }\textbf {\bibinfo {volume} {78}},\ \bibinfo
  {pages} {063421} (\bibinfo {year} {2008})}\BibitemShut {NoStop}%
\bibitem [{\citenamefont {Anderegg}\ \emph {et~al.}(2017)\citenamefont
  {Anderegg}, \citenamefont {Augenbraun}, \citenamefont {Chae}, \citenamefont
  {Hemmerling}, \citenamefont {Hutzler}, \citenamefont {Ravi}, \citenamefont
  {Collopy}, \citenamefont {Ye}, \citenamefont {Ketterle},\ and\ \citenamefont
  {Doyle}}]{Anderegg2017}%
  \BibitemOpen
  \bibfield  {author} {\bibinfo {author} {\bibfnamefont {L.}~\bibnamefont
  {Anderegg}}, \bibinfo {author} {\bibfnamefont {B.~L.}\ \bibnamefont
  {Augenbraun}}, \bibinfo {author} {\bibfnamefont {E.}~\bibnamefont {Chae}},
  \bibinfo {author} {\bibfnamefont {B.}~\bibnamefont {Hemmerling}}, \bibinfo
  {author} {\bibfnamefont {N.~R.}\ \bibnamefont {Hutzler}}, \bibinfo {author}
  {\bibfnamefont {A.}~\bibnamefont {Ravi}}, \bibinfo {author} {\bibfnamefont
  {A.}~\bibnamefont {Collopy}}, \bibinfo {author} {\bibfnamefont
  {J.}~\bibnamefont {Ye}}, \bibinfo {author} {\bibfnamefont {W.}~\bibnamefont
  {Ketterle}}, \ and\ \bibinfo {author} {\bibfnamefont {J.~M.}\ \bibnamefont
  {Doyle}},\ }\href {\doibase 10.1103/PhysRevLett.119.103201} {\bibfield
  {journal} {\bibinfo  {journal} {Phys. Rev. Lett.}\ }\textbf {\bibinfo
  {volume} {119}},\ \bibinfo {pages} {103201} (\bibinfo {year}
  {2017})}\BibitemShut {NoStop}%
\bibitem [{\citenamefont {Truppe}\ \emph {et~al.}(2017)\citenamefont {Truppe},
  \citenamefont {Williams}, \citenamefont {Hambach}, \citenamefont {Caldwell},
  \citenamefont {Hinds}, \citenamefont {Sauer},\ and\ \citenamefont
  {Tarbutt}}]{Truppe2017}%
  \BibitemOpen
  \bibfield  {author} {\bibinfo {author} {\bibfnamefont {S.}~\bibnamefont
  {Truppe}}, \bibinfo {author} {\bibfnamefont {H.~J.}\ \bibnamefont
  {Williams}}, \bibinfo {author} {\bibfnamefont {M.}~\bibnamefont {Hambach}},
  \bibinfo {author} {\bibfnamefont {L.}~\bibnamefont {Caldwell}}, \bibinfo
  {author} {\bibfnamefont {N.~J. F. E.~A.}\ \bibnamefont {Hinds}}, \bibinfo
  {author} {\bibfnamefont {B.~E.}\ \bibnamefont {Sauer}}, \ and\ \bibinfo
  {author} {\bibfnamefont {M.~R.}\ \bibnamefont {Tarbutt}},\ }\href {\doibase
  https://doi.org/10.1038/nphys4241} {\bibfield  {journal} {\bibinfo  {journal}
  {Nat. Phys.}\ }\textbf {\bibinfo {volume} {13}},\ \bibinfo {pages} {1173}
  (\bibinfo {year} {2017})}\BibitemShut {NoStop}%
\bibitem [{\citenamefont {Wang}\ \emph {et~al.}(2018)\citenamefont {Wang},
  \citenamefont {Bu}, \citenamefont {Xie}, \citenamefont {Chen},\ and\
  \citenamefont {Yan}}]{Wang2018}%
  \BibitemOpen
  \bibfield  {author} {\bibinfo {author} {\bibfnamefont {D.}~\bibnamefont
  {Wang}}, \bibinfo {author} {\bibfnamefont {W.}~\bibnamefont {Bu}}, \bibinfo
  {author} {\bibfnamefont {D.}~\bibnamefont {Xie}}, \bibinfo {author}
  {\bibfnamefont {T.}~\bibnamefont {Chen}}, \ and\ \bibinfo {author}
  {\bibfnamefont {B.}~\bibnamefont {Yan}},\ }\href {\doibase
  10.1364/JOSAB.35.001658} {\bibfield  {journal} {\bibinfo  {journal} {J. Opt.
  Soc. Am. B}\ }\textbf {\bibinfo {volume} {35}},\ \bibinfo {pages} {1658}
  (\bibinfo {year} {2018})}\BibitemShut {NoStop}%
\bibitem [{\citenamefont {Teng}\ \emph {et~al.}(2015)\citenamefont {Teng},
  \citenamefont {Disla}, \citenamefont {Dellatto}, \citenamefont {Limani},
  \citenamefont {Kaufman},\ and\ \citenamefont {Wright}}]{Teng2015}%
  \BibitemOpen
  \bibfield  {author} {\bibinfo {author} {\bibfnamefont {K.}~\bibnamefont
  {Teng}}, \bibinfo {author} {\bibfnamefont {M.}~\bibnamefont {Disla}},
  \bibinfo {author} {\bibfnamefont {J.}~\bibnamefont {Dellatto}}, \bibinfo
  {author} {\bibfnamefont {A.}~\bibnamefont {Limani}}, \bibinfo {author}
  {\bibfnamefont {B.}~\bibnamefont {Kaufman}}, \ and\ \bibinfo {author}
  {\bibfnamefont {M.~J.}\ \bibnamefont {Wright}},\ }\href {\doibase
  10.1063/1.4918731} {\bibfield  {journal} {\bibinfo  {journal} {Review of
  Scientific Instruments}\ }\textbf {\bibinfo {volume} {86}},\ \bibinfo {pages}
  {043114} (\bibinfo {year} {2015})}\BibitemShut {NoStop}%
\bibitem [{\citenamefont {Kaufman}\ \emph {et~al.}(2017)\citenamefont
  {Kaufman}, \citenamefont {Paltoo}, \citenamefont {Grogan}, \citenamefont
  {Pena}, \citenamefont {John},\ and\ \citenamefont {Wright}}]{Kaufman2017}%
  \BibitemOpen
  \bibfield  {author} {\bibinfo {author} {\bibfnamefont {B.}~\bibnamefont
  {Kaufman}}, \bibinfo {author} {\bibfnamefont {T.}~\bibnamefont {Paltoo}},
  \bibinfo {author} {\bibfnamefont {T.}~\bibnamefont {Grogan}}, \bibinfo
  {author} {\bibfnamefont {T.}~\bibnamefont {Pena}}, \bibinfo {author}
  {\bibfnamefont {J.~P.~S.}\ \bibnamefont {John}}, \ and\ \bibinfo {author}
  {\bibfnamefont {M.~J.}\ \bibnamefont {Wright}},\ }\href {\doibase
  10.1007/s00340-017-6649-3} {\bibfield  {journal} {\bibinfo  {journal}
  {Applied Physics B}\ }\textbf {\bibinfo {volume} {123}},\ \bibinfo {pages}
  {58} (\bibinfo {year} {2017})}\BibitemShut {NoStop}%
\bibitem [{\citenamefont {Yeo}\ \emph {et~al.}(2015)\citenamefont {Yeo},
  \citenamefont {Hummon}, \citenamefont {Collopy}, \citenamefont {Yan},
  \citenamefont {Hemmerling}, \citenamefont {Chae}, \citenamefont {Doyle},\
  and\ \citenamefont {Ye}}]{Yeo2015}%
  \BibitemOpen
  \bibfield  {author} {\bibinfo {author} {\bibfnamefont {M.}~\bibnamefont
  {Yeo}}, \bibinfo {author} {\bibfnamefont {M.~T.}\ \bibnamefont {Hummon}},
  \bibinfo {author} {\bibfnamefont {A.~L.}\ \bibnamefont {Collopy}}, \bibinfo
  {author} {\bibfnamefont {B.}~\bibnamefont {Yan}}, \bibinfo {author}
  {\bibfnamefont {B.}~\bibnamefont {Hemmerling}}, \bibinfo {author}
  {\bibfnamefont {E.}~\bibnamefont {Chae}}, \bibinfo {author} {\bibfnamefont
  {J.~M.}\ \bibnamefont {Doyle}}, \ and\ \bibinfo {author} {\bibfnamefont
  {J.}~\bibnamefont {Ye}},\ }\href {\doibase 10.1103/PhysRevLett.114.223003}
  {\bibfield  {journal} {\bibinfo  {journal} {Phys. Rev. Lett.}\ }\textbf
  {\bibinfo {volume} {114}},\ \bibinfo {pages} {223003} (\bibinfo {year}
  {2015})}\BibitemShut {NoStop}%
\bibitem [{\citenamefont {Brown}\ and\ \citenamefont
  {Carrington}(2003)}]{Brown2003}%
  \BibitemOpen
  \bibfield  {author} {\bibinfo {author} {\bibfnamefont {J.}~\bibnamefont
  {Brown}}\ and\ \bibinfo {author} {\bibfnamefont {A.}~\bibnamefont
  {Carrington}},\ }\href@noop {} {\emph {\bibinfo {title} {Rotational
  Spectroscopy of Diatomic Molecules}}},\ edited by\ \bibinfo {editor}
  {\bibfnamefont {A.~H.~Z.}\ \bibnamefont {Richard J.~Saykally}}\ and\ \bibinfo
  {editor} {\bibfnamefont {D.~A.}\ \bibnamefont {King}}\ (\bibinfo  {publisher}
  {Cambridge University Press},\ \bibinfo {year} {2003})\BibitemShut {NoStop}%
\end{thebibliography}%
\end{document}